\documentclass[pdflatex,sn-basic]{sn-jnl}

\usepackage{graphicx}%
\usepackage{multirow}%
\usepackage{amsmath,amssymb,amsfonts}%
\usepackage{amsthm}%
\usepackage{mathrsfs}%
\usepackage[title]{appendix}%
\usepackage{xcolor}%
\usepackage{textcomp}%
\usepackage{manyfoot}%
\usepackage{booktabs}%
\usepackage{algorithm}%
\usepackage{algorithmicx}%
\usepackage{algpseudocode}%
\usepackage{listings}%
\usepackage{comment}
\usepackage{lmodern} 

\theoremstyle{thmstyleone}%
\theoremstyle{thmstyletwo}%

\theoremstyle{thmstylethree}%

\usepackage{caption}
\usepackage{subcaption}
\usepackage{hyperref} 
\begin{document}

\title[Article Title]{Efficient Data Retrieval and Comparative Bias Analysis of Recommendation Algorithms for YouTube Shorts and Long-Form Videos}

\author*{\begin{tabular}{c}
\textbf{Selimhan Dagtas}$^{1}$ \\
\href{mailto:sedagtas@ualr.edu}{sedagtas@ualr.edu} \\
\href{https://scholar.google.com/citations?user=HUNSBlEAAAAJ}{https://scholar.google.com/citations?user=HUNSBlEAAAAJ}
\end{tabular}
\quad
\begin{tabular}{c}
\textbf{Mert Can Cakmak}$^{1}$ \\
\href{mailto:mccakmak@ualr.edu}{mccakmak@ualr.edu} \\
\href{https://scholar.google.com/citations?user=koC7AhkAAAAJ}{https://scholar.google.com/citations?user=koC7AhkAAAAJ}
\end{tabular}
\quad
\begin{tabular}{c}
\textbf{Nitin Agarwal}$^{1,2}$ \\
\href{mailto:nxagarwal@ualr.edu}{nxagarwal@ualr.edu} \\
\href{https://scholar.google.com/citations?user=V9PGXjMAAAAJ}{https://scholar.google.com/citations?user=V9PGXjMAAAAJ}
\end{tabular}
}

\affil[1]{\orgdiv{COSMOS Research Center}, \orgname{University of Arkansas - Little Rock}, \orgaddress{\state{Arkansas}, \country{USA}}}
\affil[2]{\orgdiv{International Computer Science Institute}, \orgname{University of California}, \orgaddress{\state{Berkeley}, \country{USA}}}


\abstract{The growing popularity of short-form video content, such as YouTube Shorts, has transformed user engagement on digital platforms, raising critical questions about the role of recommendation algorithms in shaping user experiences. These algorithms significantly influence content consumption, yet concerns about biases, echo chambers, and content diversity persist. This study develops an efficient data collection framework to analyze YouTube’s recommendation algorithms for both short-form and long-form videos, employing parallel computing and advanced scraping techniques to overcome limitations of YouTube’s API. The analysis uncovers distinct behavioral patterns in recommendation algorithms across the two formats, with short-form videos showing a more immediate shift toward engaging yet less diverse content compared to long-form videos. Furthermore, a novel investigation into biases in politically sensitive topics, such as the South China Sea dispute, highlights the role of these algorithms in shaping narratives and amplifying specific viewpoints. By providing actionable insights for designing equitable and transparent recommendation systems, this research underscores the importance of responsible AI practices in the evolving digital media landscape.}




\maketitle

\section{Introduction}

Recommendation algorithms have become a cornerstone of digital platforms, transforming the way users discover and interact with content. By offering personalized suggestions, these systems help mitigate the problem of information overload, enabling users to navigate vast content repositories effectively. However, their influence extends beyond utility, shaping user engagement, content consumption patterns, and even societal behaviors. As noted in \cite{deldjoo2024fairness}, recommender systems must balance their utility with fairness and transparency, ensuring equitable access to diverse content while avoiding the amplification of biases.

YouTube, one of the world’s most influential video platforms, exemplifies the power and challenges of recommendation algorithms. Historically, YouTube has been synonymous with long-form video content, offering creators opportunities for monetization through advertisements, sponsorships, and thematic partnerships. However, the emergence of YouTube Shorts has disrupted this paradigm, catering to the modern digital audience's preference for bite-sized, fast-paced content. As highlighted in \cite{rajendran2024shorts}, the rise of YouTube Shorts has significantly influenced viewer behavior, leading to a decline in engagement with long-form videos. This shift underscores the need to understand how recommendation algorithms operate across these two distinct formats and the biases that may arise as a result.

The differences between short-form and long-form content extend beyond duration. Long-form videos often provide deeper narrative and thematic exploration, fostering meaningful user engagement and discourse. In contrast, short-form videos prioritize instant gratification, often relying on visually engaging elements to capture attention. This division presents unique challenges for recommendation algorithms, as their optimization strategies may differ for each format. For instance, while long-form videos may emphasize sustained engagement metrics, short-form content often prioritizes click-through rates and immediate interactions, as discussed in \cite{wang2024trustworthy}.

The investigation of biases within these algorithms is critical. Biases can arise from multiple sources, including user interaction data, algorithmic design, and platform policies. As described in \cite{zhao2024recommender}, the integration of advanced technologies, such as large language models, offers new opportunities to address these biases by improving the interpretability and fairness of recommendations. Additionally, \cite{rajendran2024shorts} highlights the potential for recommendation systems to inadvertently prioritize certain content types, creating echo chambers that limit exposure to diverse perspectives. Understanding these dynamics is essential to ensure that algorithms serve users equitably and transparently.

Efficient data collection is a fundamental step in studying these algorithms. Despite the availability of APIs, they often fall short in providing comprehensive recommendation data, particularly for platforms like YouTube, where recommendation chains are essential for analysis. This challenge necessitates alternative methodologies, as noted in \cite{deldjoo2024fairness}, where the need for robust and scalable data collection techniques is emphasized. By enabling the large-scale analysis of recommendation patterns, such methods provide a foundation for examining the biases and behaviors embedded within these systems.

The South China Sea (SCS) adds a unique dimension to this study. This region has long been a geopolitical hotspot, with overlapping territorial claims by multiple nations and significant strategic and economic importance \cite{prc_assertiveness_scs, analyzing_scs_dispute}. The SCS’s rich natural resources and critical trade routes, which account for nearly a third of global maritime trade, make it a focal point of international relations \cite{analyzing_scs_dispute, fravel2011china}. China’s strategic use of information and digital platforms, including YouTube, to influence perceptions about its claims in the SCS underscores the importance of studying recommendation biases in this context. As explored in \cite{fravel2011china}, the dissemination of content related to territorial disputes can shape narratives and sway public opinion, raising ethical questions about the role of recommendation algorithms in amplifying such content.

This research investigates biases in YouTube’s recommendation algorithms through the lens of the SCS dispute, analyzing both short-form and long-form content. Understanding how the algorithms prioritize content related to politically charged topics, such as the SCS, is crucial for addressing concerns about algorithmic fairness and misinformation. Additionally, the efficient collection of recommendation data is critical to ensure the scalability and reliability of such analyses, especially given the limitations of YouTube’s API. To address these challenges, this study poses the following research questions:

\begin{itemize}
    \item RQ1: How can YouTube recommendations for short-form and long-form videos be collected efficiently without relying on API support?
    \item RQ2: What optimizations are necessary to enable large-scale data collection while maintaining accuracy and efficiency?
    \item RQ3: How does the recommendation algorithm exhibit biases in prioritizing user engagement, emotional responses, and toxicity within short-form and long-form videos?
    \item RQ4: How do the recommendation algorithms for short-form and long-form videos compare in their overall behavior and outcomes?
\end{itemize}

\begin{figure}
    \centering
    \includegraphics[width=0.9\linewidth]{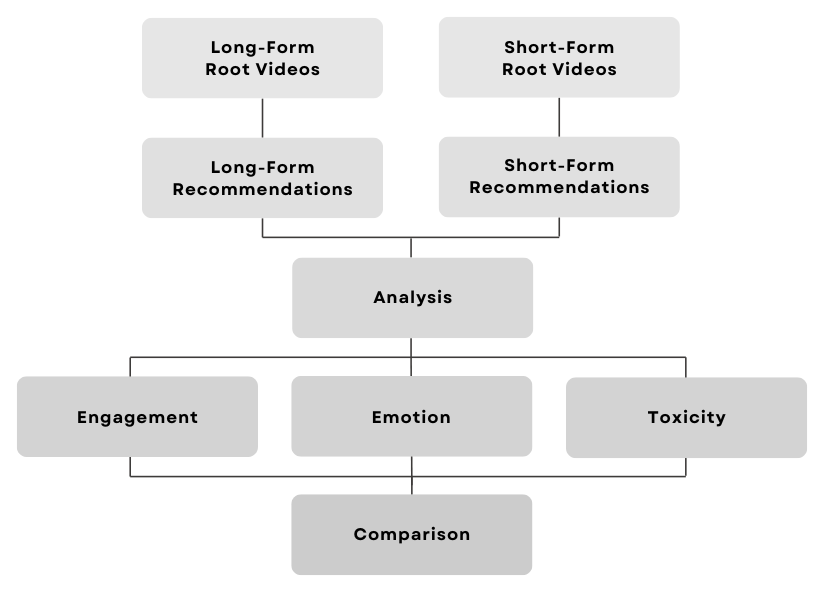}
    \caption{Research Flow Chart}
    \label{research_process_chart}
\end{figure}

By addressing these questions, this research contributes to a deeper understanding of recommendation systems, particularly their ethical implications and biases. Unlike prior studies, this work uniquely integrates a comparative analysis of short-form and long-form video algorithms, shedding light on their distinct behaviors and outcomes. Furthermore, it provides actionable insights for designing equitable and transparent recommendation algorithms, offering a framework for mitigating biases and enhancing fairness. In doing so, this study not only advances the field of recommender systems but also underscores the critical importance of responsible AI practices in shaping the future of digital content platforms. The research flow chart outlining the methodology and key processes can be observed in Figure \ref{research_process_chart}.

\section{Literature Review} \label{literature-review}

Understanding YouTube's recommendation algorithms has been a subject of significant interest due to their impact on content consumption and user engagement. Previous studies, such as \cite{zhou2010impact}, have highlighted limitations in the YouTube API, particularly its inability to directly extract recommendation data. This limitation has necessitated alternative methodologies for data collection. However, while methods for extracting recommendation data have been developed, detailed processes for efficient and accurate data collection remain underexplored.

Studies have increasingly focused on analyzing biases within YouTube's recommendation algorithms. For instance, \cite{analyzing_bias_2023, Alp2022} employed a multi-method approach to evaluate biases across recommendation depths, integrating network analysis, emotion modeling, and topic modeling to uncover systematic patterns in the algorithm's behavior. Similarly, \cite{unpacking_algorithmic_bias_2025} demonstrated the tendency of YouTube Shorts' recommendations to shift from serious, topical content to entertainment-focused themes, highlighting the algorithm's impact on user engagement and content diversity. Such findings underscore the need for more nuanced studies of algorithmic biases in both regular videos and Shorts.

Recent research has also examined YouTube Shorts as a distinct format. For example, \cite{cakmak2024unveilingbias} explored biases within the recommendations of short-form videos, revealing a strong emphasis on entertainment content. This observation aligns with \cite{violot2024shorts}, which analyzed engagement metrics and found that while Shorts achieved higher view counts, they elicited fewer comments per view, suggesting limited meaningful discourse compared to regular videos. Furthermore, \cite{bias_symbols_2025} highlighted the role of symbolic content in reinforcing cohesive but narrow recommendation networks, limiting content diversity. Recommender systems have also been accused of leading users to increasingly homogeneous content, resulting in filter bubbles and echo chambers \cite{kitchens2020understanding, gurung2024decoding} where users are isolated from diversified content and viewpoints \cite{hosanagar2014will, Treviranus2009, Ballard2022, Chaney2018}.

Metrics for evaluating recommendation algorithms have evolved alongside these studies. Emotional and moral dimensions have been identified as critical evaluative metrics in studies such as \cite{cakmak2023investigating}. Meanwhile, \cite{su2024covid} utilized toxicity as a key metric for analyzing textual content in video metadata. Engagement metrics, including views, likes, and comments, were emphasized by \cite{khan2017social} and \cite{hoiles2017engagement}, providing insights into user interaction patterns. Additionally, \cite{poudel2024beyond} analyzed thematic differences between regular videos and Shorts, noting that regular videos often encompass a broader range of topics, including science, education, and politics, whereas Shorts predominantly cater to entertainment-driven content.

Multimedia elements play a crucial role in shaping user engagement, emotional responses, and the biases inherent in YouTube’s recommendation system. \cite{cakmak2024color} demonstrated the emotional impact of color patterns in YouTube videos, revealing strong correlations between color combinations and viewer sentiment. These findings are particularly relevant when comparing YouTube Shorts and long-form videos, as the shorter format often relies heavily on visually striking elements, such as vibrant thumbnails, to capture attention quickly. Similarly, \cite{Shaik2024Characterizing} examined the role of multimedia in mobilizing social movements, showcasing how visual content fosters connectivity and collective action. This dynamic highlights the potential for biases in recommendations to amplify certain narratives or content types based on engagement metrics. Furthermore, \cite{yousefi2024examining} explored the interplay between textual, visual, and auditory elements, showing how these modalities interact to evoke user emotions. These findings provide valuable insights into how both Shorts and long-form videos are differentially prioritized in YouTube’s recommendation system, potentially reinforcing biases and shaping user behavior.

Despite these advancements, gaps remain in the comparative analysis of recommendation algorithms for regular YouTube videos and Shorts. Existing studies, such as \cite{violot2024shorts}, have highlighted the differences in engagement metrics and thematic focus, but the implications of these differences on content diversity and user behavior remain underexplored. Furthermore, while studies like \cite{bias_symbols_2025} and \cite{okeke_examining_2023} have provided insights into the biases introduced by symbolic and emotional content, their impact on the broader recommendation ecosystem requires further investigation.

This study seeks to bridge existing gaps by offering a comprehensive analysis of YouTube’s recommendation dynamics across both short-form (YouTube Shorts) and long-form videos. By addressing challenges in efficient recommendation data collection, this work provides a robust framework for studying algorithmic biases. Integrating key metrics such as engagement, emotion, and thematic focus, the study aims to deepen our understanding of how these biases influence content accessibility, user interaction, and diversity across video formats. Furthermore, this research sets itself apart by not only analyzing biases in individual formats but also drawing comparisons between Shorts and long-form videos, providing a holistic view of the recommendation ecosystem. Ultimately, this work contributes novel insights into the ethical implications of recommendation algorithms, offering a valuable foundation for future studies and practical strategies for mitigating biases in digital media.


\section{Efficient Data Collection for Recommendations} \label{data_collection}

Effective data collection is fundamental to understanding the extent of bias in YouTube's recommendation algorithm. There are many theories about the severity of the bias in YouTube's recommendation algorithm; however, these theories cannot be tested without collecting recommendation data. Since the official YouTube Data API does not support the gathering of recommendation data, a more nuanced data collection method must be developed. Our approach was to simulate a user browsing YouTube using the Python library Selenium. Selenium is a library with a wide range of automated website interactions, which was initially designed for website testing but is perfect for any application that requires the simulation of user behavior.

In the following subsections, we describe the process of using Selenium to collect recommendation data from both YouTube short-form and long-form videos. Later on, we will go through the challenges we faced with both of the video formats, but first we mention the measures we took to ensure an unbiased data collection for both formats. 

The recommendation algorithm can be influenced by a plethora of factors. This forces us to eliminate as many of those factors as possible. The clearest step is to not log in to any account when collecting data. Furthermore, the history of the browser must be empty, since the algorithm can be influenced by past watched videos. Finally the cache and cookies are cleared to further prevent the recommendation algorithm from being influenced by external factors. One factor, however, that was impossible to control for was the IP location; the internet connection has to be initiated from a certain location, and for this study we chose to connect from the United States. 

Finally, to collect root video IDs, which represent the initial videos for gathering recommendations, several methods can be employed. For example, keyword searches can be performed using the YouTube Data API, scrapers such as APIFY can be utilized, or video IDs can be obtained from certain channels with same approaches. The method chosen depends on the study’s requirements. Once the root video IDs are collected, the recommendation data collection process begins.

In this study, we have used keyword search. As shown in Table \ref{SCS-Keywords-Table}, our selected keywords for the South China Sea dispute study encapsulate the conflict's key aspects: legal rulings (``Permanent Court", ``Arbitration", ``UNCLOS"), geopolitical tensions (``China + Philippines", ``sovereignty"), and economic interests (``economic cooperation", ``natural resources"). These terms are essential to examine the intricate blend of legal, political, and economic factors in the dispute, particularly focusing on China's territorial claims and the responses of neighboring states like the Philippines. The keywords enable a comprehensive analysis, aligning with the diverse perspectives and complexities discussed in our literature review.

\begin{table}[h]
\caption{South China Sea Dispute Related Keywords}\label{SCS-Keywords-Table}%
\begin{tabular*}{\textwidth}{@{}l@{}}
\toprule
Keywords \\
\midrule
``Permanent Court" + Arbitration + South China Sea,
China + Philippines + ``South China Sea" \\
\midrule
+ sovereignty, ``South China Sea" + UNCLOS + 2016 ruling,
Philippines + ``West Philippine Sea" \\
\midrule
+ China + sovereignty, China + ``nine-dash line" + historical claims,
South China Sea + maritime \\
\midrule
`` justice" + international law,
China + Philippines + ``economic cooperation" + trade, Philippines\\
\midrule
 + ``Scarborough Shoal" + territorial dispute,
``West Philippine Sea" + environmental concerns \\
\midrule
 + coral reefs, China + ``Maritime Silk Road" + regional influence,
South China Sea + ASEAN + \\
\midrule
 regional stability,
Philippines + China + ``mutual benefit" + diplomacy,
China + ``artificial  \\ 
\midrule
islands" + military presence,
South China Sea + ``international arbitration" + dispute resolution,\\ 
\midrule
China + Philippines + ``joint exploration" + resources. \\
\botrule
\end{tabular*}
\end{table}

\subsection{YouTube Short-Form Video Recommendation Collection}

Before the YouTube Shorts recommendation scraping process begins, the initial videos for which we want to collect recommendations, referred to as root videos, are imported into the program from a predefined CSV file containing their video IDs. The code can be modified to import data from an Excel or TXT file as well. Once the root videos are imported, the scraping process begins.

The process for short-form videos started with initializing the web driver using Selenium and opening the browser and as well as opening the root videos in the browser. The opening of the browser was included in the main scraping loop to ensure that the program opened a new browser with a clear cache and history for each root video. Eliminating possible bias-inducing factors like cache and history guaranteed that any bias seen in the recommended videos was purely from YouTube’s recommendation algorithm.

Using the link format, ``https://www.youtube.com/shorts/\textit{VIDEO\_ID}'', the root videos are opened. Once the root short-form video was opened, YouTube automatically loaded in recommendations under the opened video. Normally, the user would scroll down to start watching the next video recommended by YouTube. In this case, the ‘user' is the program, and scrolling down can be implemented in Selenium in a few different ways.

\begin{figure}
    \centering
    \includegraphics[width=0.8\linewidth]{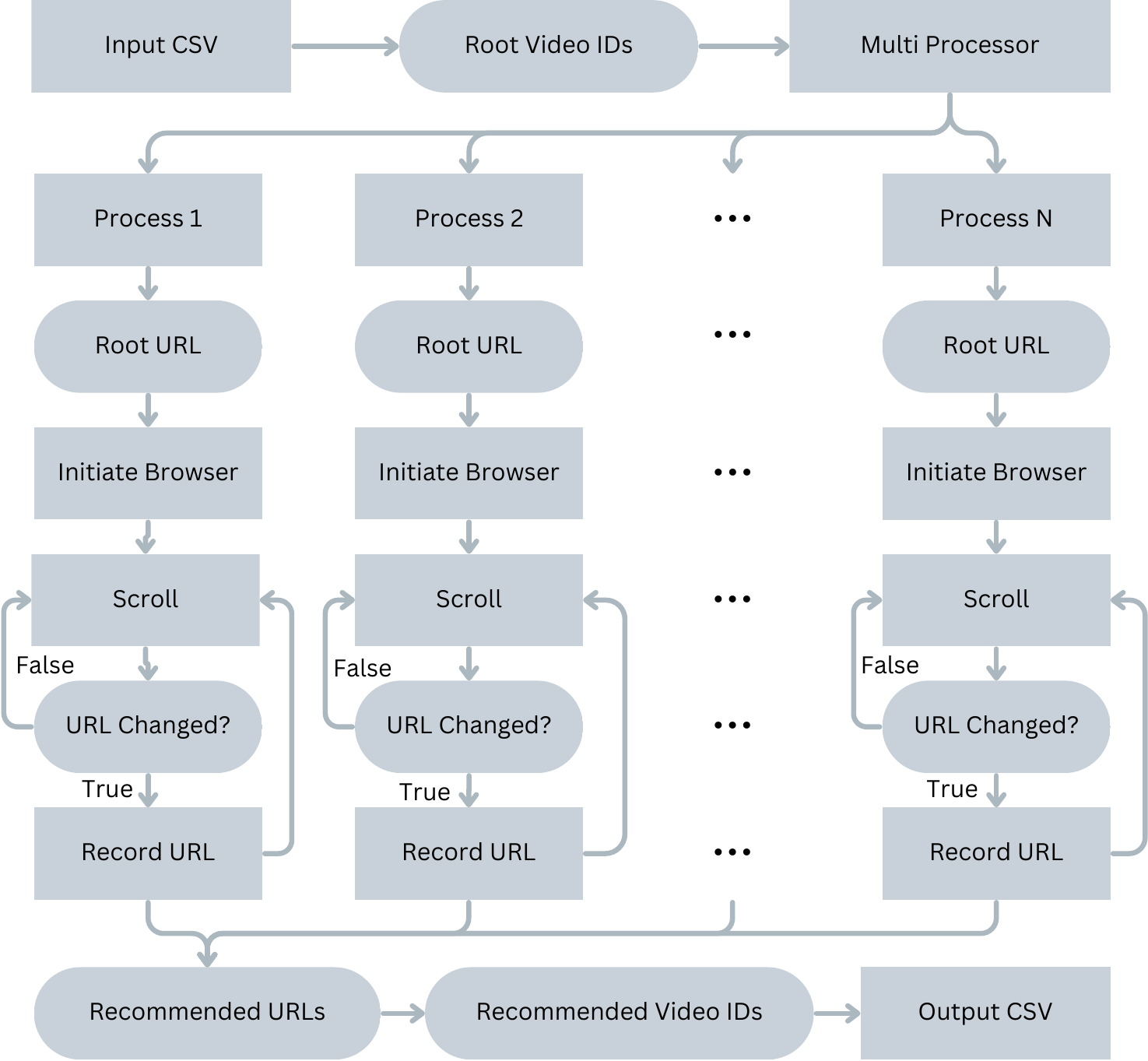}
    \caption{Flow Chart of YouTube Shorts Scraping}
    \label{shorts_scraping_flow}
\end{figure}

As represented in Figure \ref{shorts_scraping_flow}, the recommendation collection for short-form YouTube videos involves evenly distributing the root videos into multiple parallel processes that scroll through the videos as they collect the URLs of the recommended videos. This methodology also includes a fail safe to ensure the scroll input was registered by the browser by checking if the URL has in fact changed, meaning a new video is being played.

\subsubsection{Scrolling using Selenium}

To collect recommendations for short-form videos, it was necessary to simulate scrolling behavior similar to that of a real user. One method uses the buttons on the bottom right that the user can click to scroll up or down. To implement this, first the button element needs to be identified and then clicked. This is not ideal since large social media platforms occasionally change the tags and classes of elements, and there are more robust methods of scrolling down. One alternative uses the scroll wheel input using Selenium. This method works regardless of any tag or class change YouTube might implement on any part of the website. However, using the scroll wheel introduces a new problem stemming from the fact that to scroll down, the program needs a specified amount to scroll down. If selenium’s scroll down function is given too low a value, YouTube will not scroll to the next video, and if the function is given too high a value, YouTube will skip one or more recommended videos. The final and best alternative uses the arrow keys or the page up and down keys. Selenium still requires an element to attach the “send keys" function to. The most stable and robust element to do this with is the “html" element that encompasses the whole page and cannot be removed. Using one of these key pairs makes the program precise in its scrolling and adaptable against changes YouTube might implement in its HTML. For simplicity, the down arrow key was used in Selenium in this way.

\subsubsection{Wait Time Optimization}

Once the page is scrolled and a new video appears on the screen, the next step is to capture the video ID of the newly displayed content. The simplest and most robust way to do this is to use the URL of the page which looks like in this format ``https://www.youtube.com/shorts/\textit{VIDEO\_ID}''. When the program scrolls down, and the next recommended video appears. The URL is updated to contain the video ID of the new video. This makes it easy to collect the ID of the new video. Furthermore, the wait time when scrolling between two consecutive videos is important. This can be done by remembering the old URL or video ID, and continuously checking the URL to detect the new URL as soon as it has changed. This checking mechanism replaces any arbitrary wait time that would have otherwise been implemented.  

To improve on this basic recommendation scraping method and to simulate an actual user interaction in a more accurate fashion, a wait time was introduced to allow the video to play a certain amount. This was done because a user generally does not watch a short-form video for a few milliseconds and immediately jump to the next video. Users generally watch a video for either its entire duration or a significant fraction of the video. Furthermore, our hypothesis was that recommendations would be affected and altered depending on the user's watch time (in seconds), since Shorts videos' recommendations are not immediately visible like regular videos' are. Therefore, this is also important to consider while collecting recommendations to ensure accurate recommendations. To simulate this, a wait function was implemented between collecting the video ID and scrolling to the next video. This wait time can be defined as needed, and for our case the wait times of 3 seconds, 15 seconds, and 60 seconds were chosen.

\subsubsection{Multiprocessing}

\begin{figure}[h]
    \centering
    \includegraphics[width=1\textwidth]{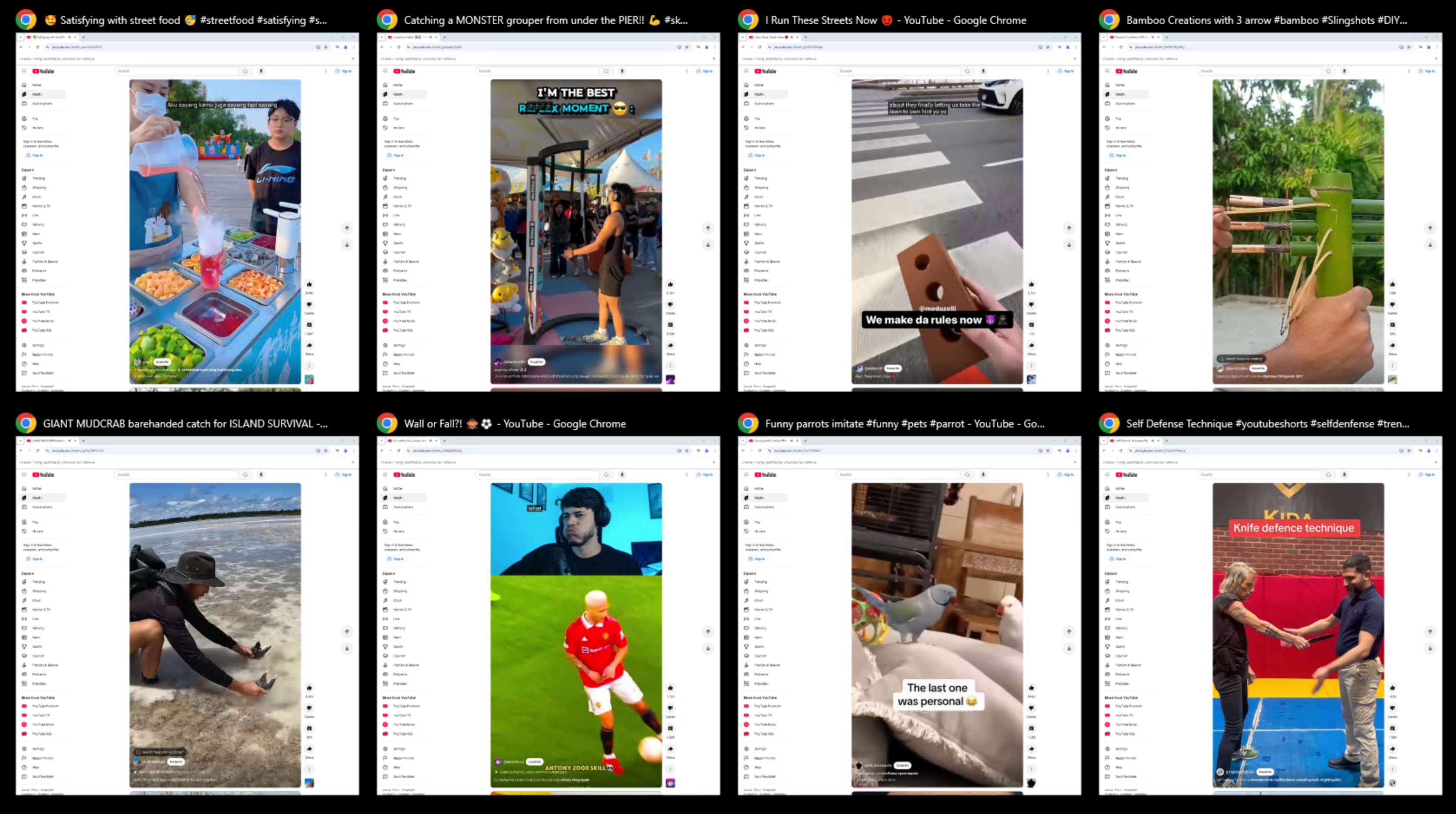}
    \caption{YouTube Shorts Collection with Multiprocessing}
    \label{short_collection}
\end{figure}

Introducing wait times drastically increases the collection time. For example, collecting recommendations for 2,000 short-form videos with a 10 recommendation depth and 60 seconds wait time would theoretically take 13.9 days to complete without any optimization. Therefore, this straightforward method was not a viable solution to collecting any meaningful amount of data in any reasonable amount of time. Hence, parallel processing was implemented using Python’s ``multiprocessing'' library to decrease the total duration of recommendation collection. This theoretically decreases the time it collects the data by a magnitude of the number of processes used. This was accomplished by first splitting the root videos for recommendation scraping and then starting each on a separate process. Each browser instance was run in a separate process to enable parallel execution. Multiple browser instances are opened using each multiprocessing pool, as illustrated in Figure \ref{short_collection}, which provides an example of the YouTube Shorts recommendation collection process. The number of processes used depends on the capabilities of the computer used. This has the potential to increase the speed of recommendation collection by ten- or twenty-fold. However, there is one more optimization that can be done to get the most amount of recommendation data in the shortest amount of time.

\subsubsection{Asynchronous Scrolling}

Within the individual data collection process, the computational load of scrolling to the next recommended video is exceedingly higher than that of waiting and watching the video. Traditionally, if 10 processes are being used to collect several recommendations, all of the processes start at the same time. This results in a synchronization of the scroll timings for all the processes. This is problematic because it causes a spike in the computational demand every 60 seconds or however long the wait time is defined. A better way to approach this is to distribute the scrolling timings of the processes evenly throughout the wait time. This creates a much smoother computational load with many small spikes within the defined wait time. Hence, the maximum computational load with a given number of processes is decreased, and the number of processes allowed is increased, without crashes or glitches, for a given system. 

\subsection{YouTube Long-Form Video Recommendation Collection}

The first step in collecting recommendations for long-form YouTube videos was to open the root video in YouTube with this URL format ``https://www.youtube.com/watch?v=\textit{VIDEO\_ID}''. Before collecting the recommendations, it was critical that the program ensured the page had completely loaded in, in order to avoid errors; however, it was equally important not to wait an excessive amount of time due to the exponential consequences. This was ensured using a three-step process. Firstly, within the Selenium WebDriver class, there is an option variable called \textit{implicitly\_wait}. Setting this option to a set number of seconds caused Selenium to wait that set amount of time before returning with an error if the element being called had not been loaded yet. Furthermore, the program was paused while searching for the title element, one of the main components of the YouTube video page, and continued only after the title element had been loaded in. The final step was an additional one-second pause for good measure and to allow other less essential components of the page to load. 

\begin{figure}
    \centering
    \includegraphics[width=0.9\linewidth]{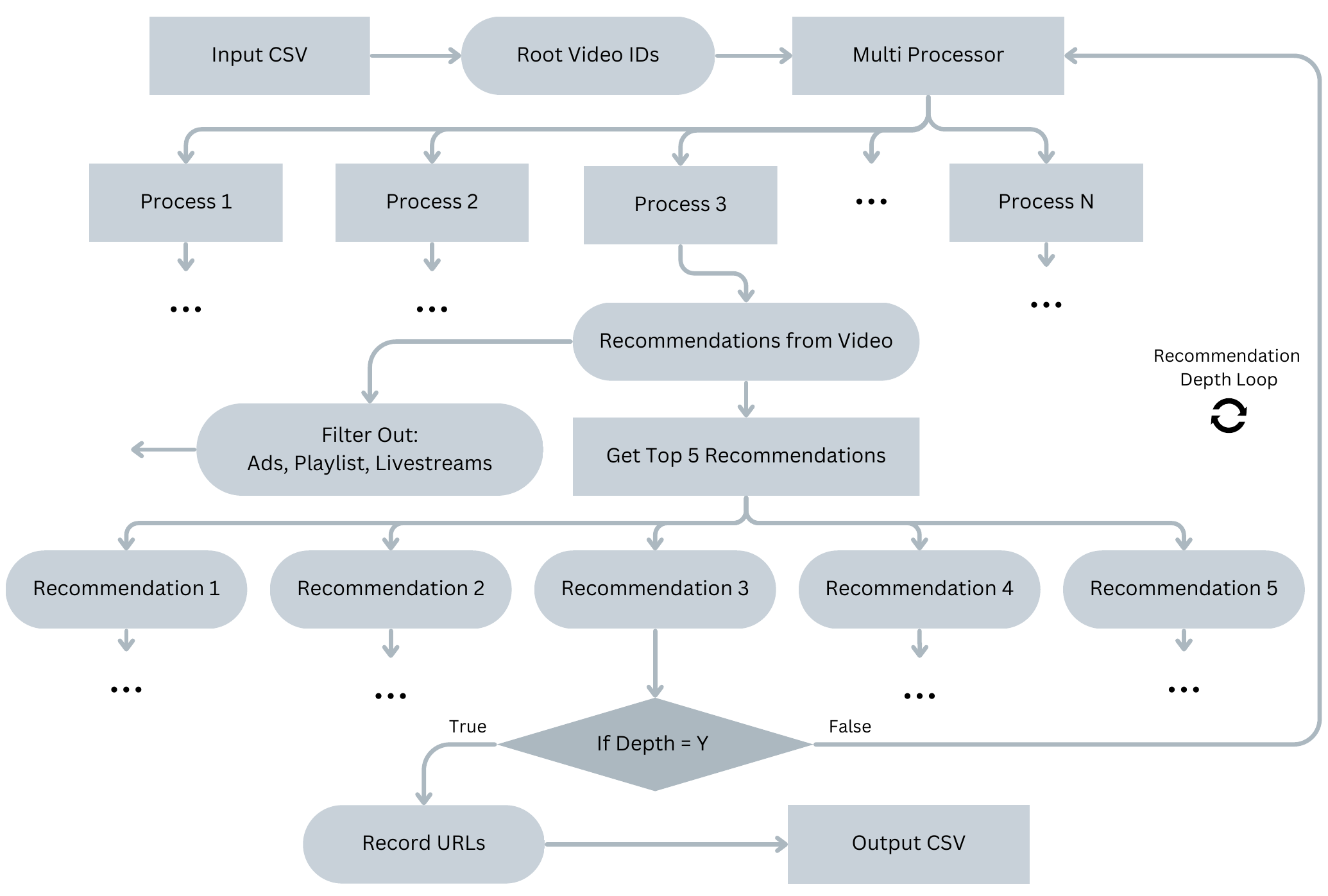}
    \caption{Flow Chart of YouTube Long-form Video Scraping}
    \label{long_form_scraping_flow}
\end{figure}

As depicted in Figure \ref{long_form_scraping_flow} the recommendation collection for long-form videos starts off, similar to short-form recommendations, by separating the root videos evenly between N number of processes. Then the top five recommendations for each root video is collected. This process is looped for a predefined number of desired depth, Y.

\subsubsection{Collection of URL Tags}

\begin{figure}[ht]
    \centering
    \includegraphics[width=0.85\textwidth]{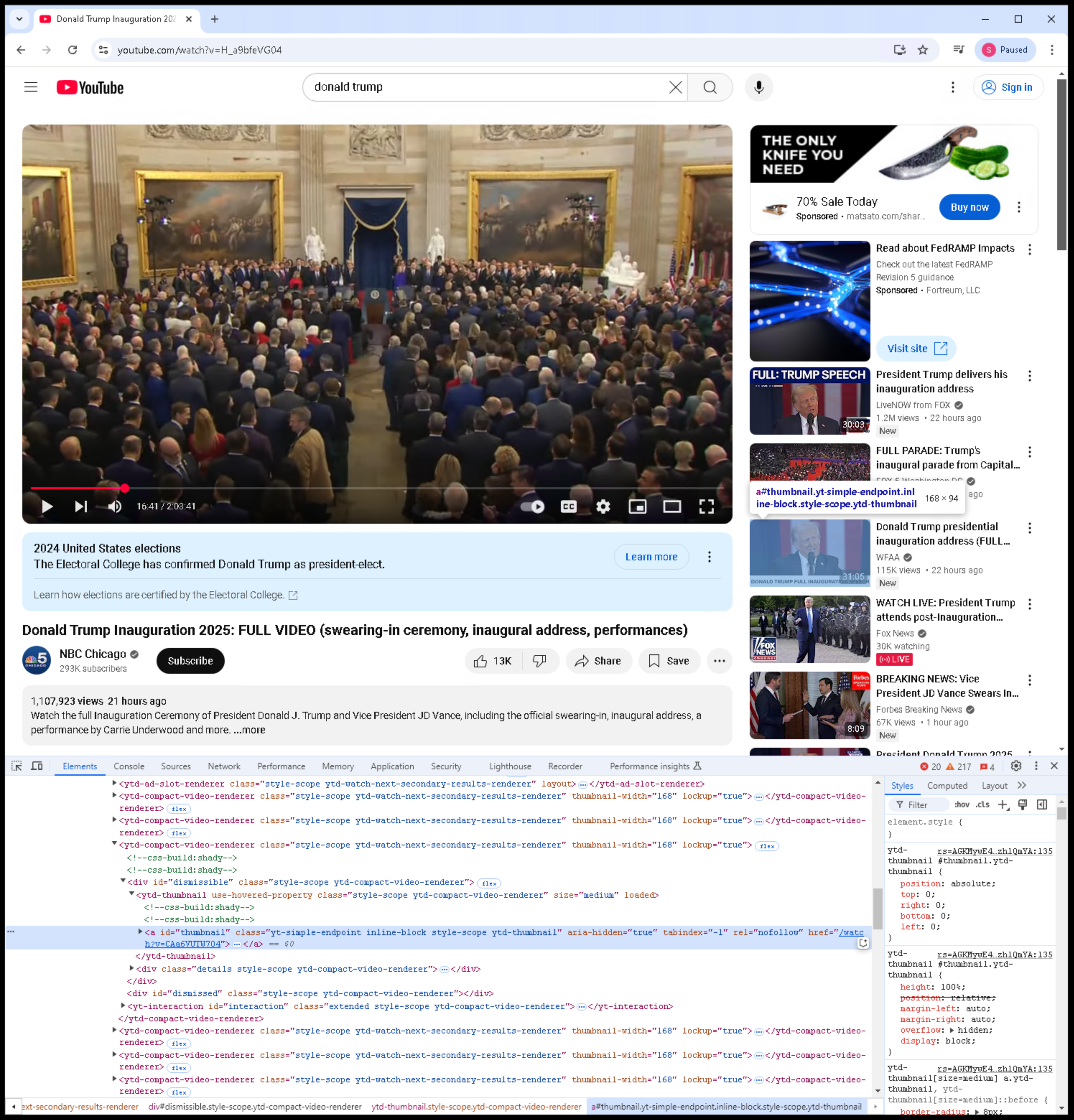}
    \caption{Engagement Metrics of Long-Form Videos}
    \label{thumbnail_id_ex}
\end{figure}

Once the page had fully loaded in, the recommendations could be scraped. On YouTube, the recommendations are listed in a column on the right side of the page. Each recommended video is represented by a video thumbnail block and a description block that houses the title, the channel, number of views, and time passed since the video was uploaded. Block examples has been showed in below and in Figure \ref{thumbnail_id_ex}:

\vspace{4mm}

Description Block Example:
\vspace{2mm}
\begin{verbatim}
<a class="yt-simple-endpoint style-scope ytd-compact-video-renderer" 
rel="nofollow" href="/watch?v=uWWVNq5GHp4"> … </a>
\end{verbatim}

\vspace{4mm}

Thumbnail Block Example:
\vspace{2mm}
\begin{verbatim}
<a id="thumbnail" class="yt-simple-endpoint inline-block style-scope 
ytd-thumbnail" aria-hidden="true" tabindex="-1" rel="nofollow" href=
"/watch?v=uWWVNq5GHp4"> … </a>
\end{verbatim}

\vspace{4mm}

Both block elements have the video ID of the recommended video they represent. However, the thumbnail element has an ID attached to the tag named “thumbnail” whereas the description element only has a class. This class could be used to collect the recommended videos but this same class is currently used for the title and description of ads and video playlists as well. Furthermore, the class defined consists of character and numbers which means it is subject to change in the future. Therefore, using the ID of a tag for locating it is significantly more reliable than using the class identifier. This is why the thumbnail block was chosen as the focus for collecting the recommended videos.

For optimization, we aimed to investigate the scrolling functionality of browsers as well, in order to see whether scrolling had an effect on speed optimization. For this we did two experiments, one being synchronous and the other asynchronous. In synchronous, the browsers scrolled at the same time, causing a spike on CPU usage. In the asynchronous, the scrolling happened at different times, since we waited intentionally during scrolls so that they did not scroll at the same time.

The next step was to filter the elements and reduce the size of the recommendation video IDs list to the required length, which, as mentioned above, was recommended to be set as five. There were three possible non-video elements that could be present in the recommendation’s column: ads, video playlists, and live streams.

\subsubsection{Filtering}

Using the thumbnail element already dodged the ads, since the div section for advertisements does not contain a block with an ID equal to the thumbnail. Furthermore, the video playlists were filtered by the fact that the link format is different, which was another reason thumbnail elements were focused on. Regular videos, by comparison, have a hyperlink tag attached with a string that starts with “https://www.youtube.com/watch?v=”. Filtering out the live streams was a more intricate problem, because the HTML for live streams is very similar to the HTML for regular videos. There were a few indicators in terms of the HTML to differentiate the two. The most practical and robust indicator was the ‘is-live-video' property on the parent element of a thumbnail. 

\vspace{4mm}

Is Live Video Tag Example:
\vspace{2mm}
\begin{verbatim}
<ytd-thumbnail use-hovered-property="" class="style-scope ytd-compact-
video-renderer" size="medium" is-live-video="" loaded=""> … </ytd-
thumbnail>
\end{verbatim}

\vspace{4mm}

Checking for that property and then filtering out the live videos produced a remaining list of video links for the initial recommendations loaded, typically with a length between 10 and 15. Note that if the number of recommendations to be collected per video exceeds that range, it is highly recommended to implement a scrolling function after giving the page time to load and before collecting the recommendations, to ensure YouTube adds more videos to the recommended column. The list is further simplified by extracting the video IDs from the video URLs by removing the initial “https://www.youtube.com/watch?v=” from the strings. Finally, the list is reduced to the desired size, with a default of 5, which is safely in the no need to scroll range, if not specified by the user.

This method above allowed the program to scrape the recommendations for YouTube videos in a robust and customizable manner. Sometimes, the YouTube page had an error loading, and the page would crash. In this situation, the only solution was to refresh the page or to restart the scraping process. Therefore, the scraping function above was put into a loop that called this function repeatedly until the returned string of recommended video IDs was the required length. This was further put into a loop to allow the scraper to collect recommendations to a desired depth. The program scraped the first depth, then the second, and so on. This created a logistical problem since each depth had an exponentially greater number of videos to collect than the previous depth. This exponential growth of demand increased the time it would take to collect the required data to days, sometimes weeks. This was not an acceptable situation and had to be optimized.

\subsubsection{Multiprocessing}

To address the challenge of exponential growth in recommendation collection, we implemented multiprocessing. This approach significantly reduced the time required for data scraping by enabling the program to run multiple processes concurrently. By distributing the workload across multiple CPU cores, multiprocessing provided an efficient solution to handle the vast number of recommendations generated as we recursively explored recommendations of recommendations.

We utilized Python’s multiprocessing library to achieve this parallelization. Specifically, the Pool class was employed to manage the processes, allowing us to execute tasks in parallel and aggregate their results efficiently. Each process handled a subset of root videos and their subsequent recommendation chains. For each process, a new browser instance was launched using Selenium to ensure isolation and independent data collection, preventing interference between processes. The results from all processes were dynamically collected and combined into a single output.

The aggregated data was then structured into a Pandas DataFrame for seamless post-processing and analysis. This DataFrame not only facilitated the organization of the collected data but also allowed for straightforward export to a CSV file. The use of multiprocessing not only mitigated the computational bottleneck but also ensured scalability for handling the extensive recommendation network, making the data collection process more robust and time-efficient.


\section{Analysis}

In this section, we overview the methods used to analyze the recommendation algorithms using the data collected. As mentioned previously, the recommendation algorithms for both short- and long-form videos have an immense impact on the content that users are exposed to. Furthermore, these algorithms may have tendencies to show one type of content over another. This is called recommendation bias. This bias in the recommendation algorithm might prefer to show the user one video or a handful of videos from a sea of seemingly infinite content depending on a verity of variables. Hence, to study and understand the recommendation algorithms better, it is required to analyze the recommendations across multiple metrics. The metrics chosen to be analyzed in this research are engagement, emotion, and toxicity. 

\subsection{Engagement}

Analyzing engagement metrics is essential for understanding biases in video recommendation systems. Metrics like views, likes, and comments serve as indicators of a video's popularity and audience interaction. These metrics may reveal whether recommendation algorithms disproportionately favor popular content, potentially creating a bias in the recommendations.

Our analysis was based on the hypothesis that recommendation algorithms might favor videos with higher engagement metrics as users delve deeper into content recommendations. This could create a feedback loop where popular videos gain even more visibility, overshadowing less-viewed content, regardless of its relevance or quality. For this study, we identified popular videos as those with high view counts, substantial likes, and significant comment activity.

To investigate this, we tracked the engagement metrics of videos across different recommendation depths. This method enabled us to identify patterns in how the algorithm prioritized content based on engagement. If videos with higher engagement consistently dominated recommendations, it would suggest a bias toward popular content.

Understanding how engagement metrics influence recommendations is vital, as such biases may limit the visibility of newer or highly relevant content. This could narrow the range of ideas and perspectives presented to viewers, reducing content diversity. By analyzing these metrics, we aim to uncover potential biases and gain a deeper understanding of the factors shaping recommendation systems on digital platforms.

\subsection{Emotion}

The emotional dynamics of social media content can heavily influence user engagement and can make or break a recommendation algorithm. The content being consumed by a user can have a noteworthy impact on the emotional state of the user. Depending on their emotions, the users may choose to engage with the content, engage with different content on the platform, or even leave the platform altogether. Therefore, the recommendation algorithm is intrinsically incentivized to recommend users the type of content that will keep them emotionally invested. Although measuring the user's average emotional state after watching a piece of content is not logistically possible, the metadata of the videos can be used in combination with emotional analysis models to detect the emotional conveyance of the video.  

The emotional expressions of the content can be measured by looking at a few qualities of the metadata extracted from videos. The first and the most obvious quality of a video to look at with regards to its emotional tendencies is its title. However, the amount of information that can be drawn from a video's title is clearly limited. Therefore the next piece of data to look at is the description, in which the author of a video puts the video's general idea or summary. This can be a much more comprehensive text-based representation in regards to the emotionality of the video itself. Furthermore, the transcription of a video's audio can be used to apply to the same emotional analysis to understand the video's emotional contents. Finally, the comments under the video can be collected and used in a similar analysis to demonstrate the emotion inspired in the users by the content. 

All of the video data mentioned can be pulled using the YouTube Data API except for the transcripts. For collecting the transcriptions of YouTube videos, there have been a few successful methodological approaches such as described in \cite{CakmakAdopting2023} and in \cite{cakmak2024high-speed}.  The transcript collection in this research mostly used the YouTube Transcript API, since this API had transcripts available for most of the videos. However, for the videos that did not have transcripts available, the video's audio was downloaded and transformed into text using OpenAI's Whisper model. These methods were inspired by and followed according to \cite{cakmak2024bias}.  

After the data for the text-based attributes of the video had been extracted, the next and final step was to use an emotional analysis model to detect what emotions were present in the various data collected. There were a few current models we considered that could be viable for emotion detection. The first option was to use a GPT-based model such as GPT4. Although GPT4 is known for being competent in a wide range of scenarios as a ‘jack of all trades,' it is not specialized in emotional analysis as shown in \cite{kocon2023chatgpt}, hence, there were more effective and specialized tools to conduct this analysis. The BERT model was another option that was considered. BERT would've been a good option to analyze the emotion in the text collected if a more efficient variant did not exist, namely RoBERTa as shown in \cite{liu1907roberta}. RoBERTa has also been further improved by DistilRoBERTa. 

Finally, in this analysis we decided to use a refined version of DistilRoBERTa, downloaded from Hugging Face, specifically trained on identifying a collection of emotions: anger, disgust, fear, joy, neutral, sadness, and surprise. Using this model we were able to analyze the pieces of data that we extracted from the videos and uncover the underlying emotional levels of the videos recommended by the algorithm.

\subsection{Toxicity}

In exploring biases in video recommendations, measuring toxicity is just as important as assessing emotional content. Toxicity refers to content that is rude, disrespectful, or unreasonable, and it can significantly impact how people interact with online media. Toxic elements in videos may subtly influence viewer preferences and behaviors, potentially affecting how recommendation algorithms function. Including toxicity as a factor in our analysis was therefore essential for gaining a comprehensive understanding of the biases that shape user engagement and content consumption patterns.

To assess toxicity in video content, we used the Detoxify model developed by Hanu (2020). Detoxify is a cutting-edge machine learning tool designed to detect and quantify toxic behavior in text. For this study, we selected the “unbiased” version of the model, which minimizes biases related to factors such as gender, race, or ideology. Built on a transformer-based architecture like BERT, the unbiased Detoxify model is effective at context-sensitive analysis and has been trained on a diverse dataset, allowing it to identify and score a range of toxic behaviors, including insults, threats, and hate speech.

We applied the Detoxify model to various textual elements of videos, such as titles, descriptions, transcriptions, and user comments. This process generated toxicity scores for each element, providing a quantifiable measure of toxic content within recommended videos. These scores allowed us to systematically evaluate the prevalence and severity of toxicity in video recommendations and to examine its potential influence on viewer behavior and preferences. This analysis was critical for understanding how toxic content affects the users' experience and the biases in recommendation algorithm.


\section{Results} \label{results}

This results section includes both the improvement metrics for the optimizations implemented in the recommendation data collection and the analysis conducted on the recommendations themselves.

\subsection {Data Collection Optimizations}

The data collection, as mentioned in Section \ref{data_collection}, was separated into two parts: the collection of short-form videos and long-form videos. The process of collecting these two type of videos was almost completely different. The optimizations implemented were also different and resulted in different time gains for varying reasons. Therefore, the results of these data collection optimizations were also separated in the two sections below.

\subsubsection{Collecting Short-Form YouTube Recommendations}

The optimizations for the data collection of YouTube Shorts were theoretically decreasing, but we decided to test and measure exactly how much time was saved with these improvements. This data collection tool was developed for large scale analyses with thousands of videos. To test the time saved, we would have to run the data collection without the improvements implemented. However, if we had run a realistic data collection scenario such as with 100 root videos with 50 depths of recommendations and a 60-second watch time, it would take months for the data to be collected. Hence, we decided to test the optimizations with a much smaller data sample of 100 root videos and 5 depths of recommendations with a 60-second watch time. This ensured that the measurements would not take an unreasonable amount of time.

The collection of short-form videos on YouTube involved a number of different optimizations that affected the results. Some of the methodological optimizations mentioned in Section \ref{data_collection} made sure time was not wasted between videos and ensured the collection of accurate data. These accuracy improvements cannot be empirically measured and are not included in the following results. However, the optimizations that increase the speed of the data collection such as using multiprocessing and asynchronous scrolling had an impact that could be measured. 

\begin{table}[h]
\caption{Data Collection Times for Short-Form YouTube Videos}\label{Shorts-Collection_Time}%
\begin{tabular}{@{}ccc@{}}

\textbf{Number of Processes} & \textbf{Time [Asynchronized]} & \textbf{Time [Synchronized]}  \\
\midrule
1   &  32,594 & 32,594 \\

5   &  6,764  & 6,723  \\

10  &  3,417  & 3,371  \\

15  &  2,707  & 2,699  \\

20  &  2,046  & 2,037  \\

25  &  1,406  & 1,407  \\

30  &  1,387  & 1,388  \\

35  &  1,379  & \textbf{1,383}  \\

\textbf{40}  &  \textbf{1,374}  & 1,397  \\

\end{tabular}

\end{table}

Looking at Table \ref{Shorts-Collection_Time}, it is immediately evident that using a higher number of processes has a direct correlation to the data collection speed, reducing the time it takes for the collection of the data. 

Furthermore, we can see that, initially, the data collection with synchronized scrolling was finishing faster than the data collection with asynchronized scrolling. This was due to the initial wait time to offset the processes to ensure that the scrolling was initiated evenly. This optimization was to prevent the computer hardware from experiencing computational load spikes so for low processes, it was not expected to affect the data collection times that much. However, when looking at higher number of processes, there was a clear improvement between the two. The improvement may seem insignificant with this small a sample of data collection, but with large scale data collection requests, it becomes more significant.

The initial value representing the time that it took to collect the data with a single process was 32,594 seconds. This was done with a single run for both columns since there was no additional processes to synchronize the scrolling to. Furthermore, we can see that the time it took to collect the data was higher that the theorized time. The time it should theoretically take is the number of root videos times the depth of the recommendations collected times the wait time which is calculated as 100 x 5 x 60 = 30,000. The extra two and a half thousand seconds come from the fact that the opening and closing of the browser instances take a small amount of time. Therefore, for the experiments proceeding, we expected the actual time to be at least about 7\% higher than the theorized time.

\subsubsection{Collecting Long-Form YouTube Recommendations}

In this section, we analyze the data collection optimizations made for the collection of recommendations for long-form YouTube videos. Contrary to the YouTube Shorts recommendation collection, the collection of regular YouTube video recommendations was improved more with data accuracy increase optimizations than with time decrease optimizations. Therefore, the results below show only the time improvements made by implementing multiprocessing. 

Similar to the testing and evaluation process for collecting short-form videos, a small test sample was utilized to measure improvements in this context. Specifically, the test sample consisted of 50 root videos, with five recommendations collected per video and a recommendation depth of three. The recommendations for regular videos exhibited exponential growth. For instance, starting with a single video, the top five recommendations were collected. Subsequently, the top five recommendations for each of those recommended videos were also gathered, and this process continued recursively. To ensure the feasibility of data collection within a manageable time-frame, it was crucial to limit the recommendation depth in the test sample.

Furthermore, the accuracy of the data collection tool for long-form YouTube videos was also optimized and this was done by eliminating the collection of video playlists, live-streams, advertisements, and anything that was not a regular YouTube video recommended to the user by the recommendation algorithm. Similarly, these improvements con not be measured empirically, hence will not show in Table \ref{Regular-Collection-Time}.

\begin{table}[h]
\caption{Data Collection Times for Long-Form YouTube Videos}\label{Regular-Collection-Time}%
\begin{tabular}{@{}cc@{}}
\textbf{Number of Processes} & \textbf{Time (seconds)}  \\
\midrule
1   &  6404.8 \\

5   &  1323.2  \\

10  &  723.9  \\

15  &  540.7  \\

20  &  480.4  \\

25  &  454.0  \\

30  &  476.3  \\

35  &  472.5  \\

\textbf{40}  &  \textbf{471.9}  \\

\end{tabular}

\end{table}

Looking at Table \ref{Regular-Collection-Time}, it is immediately evident that the decrease in the time taken for the data collection was almost proportional to the number of processes used. There could be some errors or some micro-glitches during the loading of the page that caused slight variances but the general trend was that using more processes resulted in faster data collection. This statement, however, starts to break down when increasing the number of processes, which pushes the limits of the pages loading. As mentioned in Section \ref{data_collection}, the data was collected as soon as the page is loaded, the bottleneck most likely stemmed from the network connection and, hence, the number of pages that can be loaded in a given amount of time. When that limit was reached, increasing the number of processes resulted in diminishing returns. 

\subsection{Engagement Analysis}

The first metric analyzed in the recommendations for YouTube's algorithm was the engagement the users had with the video content. There were three categories of engagements, and these categories were the same across short-form and long-form content. The first one was the viewership. The view count, as the name suggests, represented the number of times the video was viewed by any user; to clarify, repeat views by the same user were also included. The second type of engagement was video likes, where users click a thumbs up icon to show that they like and enjoy the content they are watching. The third, final way a user can engage with videos on YouTube, whether they are short form or long form, is by posting a comment under the video. These comments can be seen by the account that posted the video and any other users who are watching the video and have navigated to the comments section. This type of engagement shows that the users find the video interesting and would like to add some information or voice their own opinions on the subject. 

Before diving into the results of the analysis, it is important to note that the number of views will always be the highest measured metric on a video. Generally, they are followed second by the number of likes that the video received and finally by the number of comments left under the video. Since the discrepancies of views, likes, and comments on videos can vary drastically, we decided to use a logarithmic scale to make the visualizations more readable. Furthermore, using a logarithmic scale allowed us to put all three engagement measuring metrics into one visualization, making it easier to compare the trends in each side by side. 

\begin{figure}[h]
    \centering
    \includegraphics[width=0.78\textwidth]{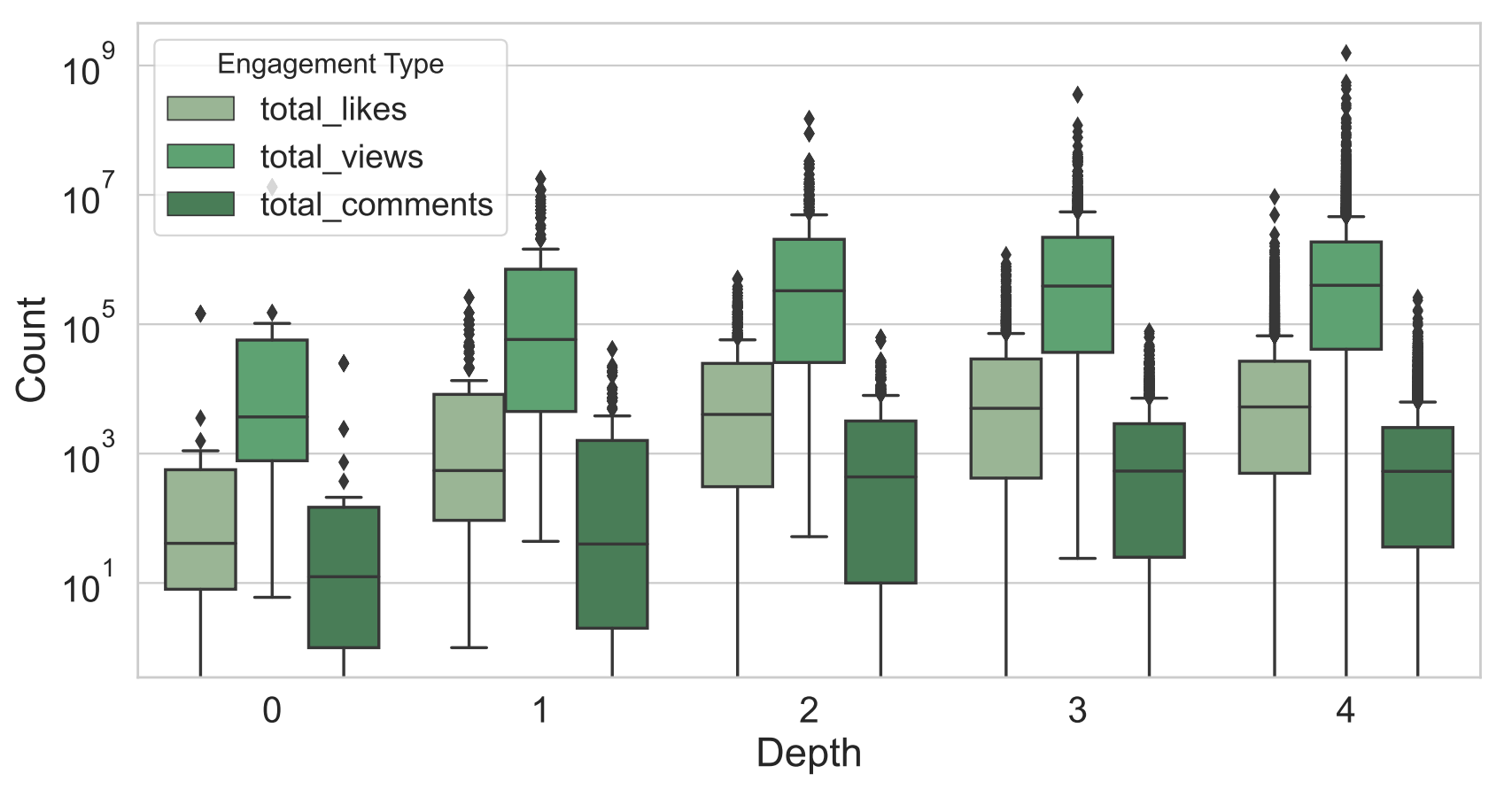}
    \caption{Engagement Metrics of Long-Form Videos}
    \label{engagement-long}
\end{figure}

When examining Figure \ref{engagement-long}, the first thing to point out is that there was a general increasing trend in all three metrics. This increase in engagement was gradual and tapered off slightly going from depth 2 to 3. This clearly shows that YouTube's recommendation algorithm for long-form videos recommends and pushes users towards videos of higher engagement. 

There is another trend that increased gradually as the depths increased, and that was the variance in the data. In the first depth, the data points were grouped close to each other and with little variance. However, as the depths increased, the data became more variant, and the box plots' minimum and maximum lines show this. This occurrence was largely caused by the fact that the higher depths have exponentially more videos than the previous ones. Hence, it is expected that a depth with more videos will have higher variance in these engagement metrics. This is evident especially when looking at some of the outliers in depth 4 that reach more than a billion views. 

Another thing to note about the results in Figure \ref{engagement-long} is the fact that the minimum values of the comments were always zero. This happens when the authors of the videos choose the option to disable the comments. Especially in political topics, this is a common occurrence. 

\newpage

\begin{figure}[h]
     \centering
     \captionsetup[subfigure]{justification=centering}
     \begin{subfigure}[b]{0.45\textwidth}
         \centering
         \includegraphics[width=\textwidth]{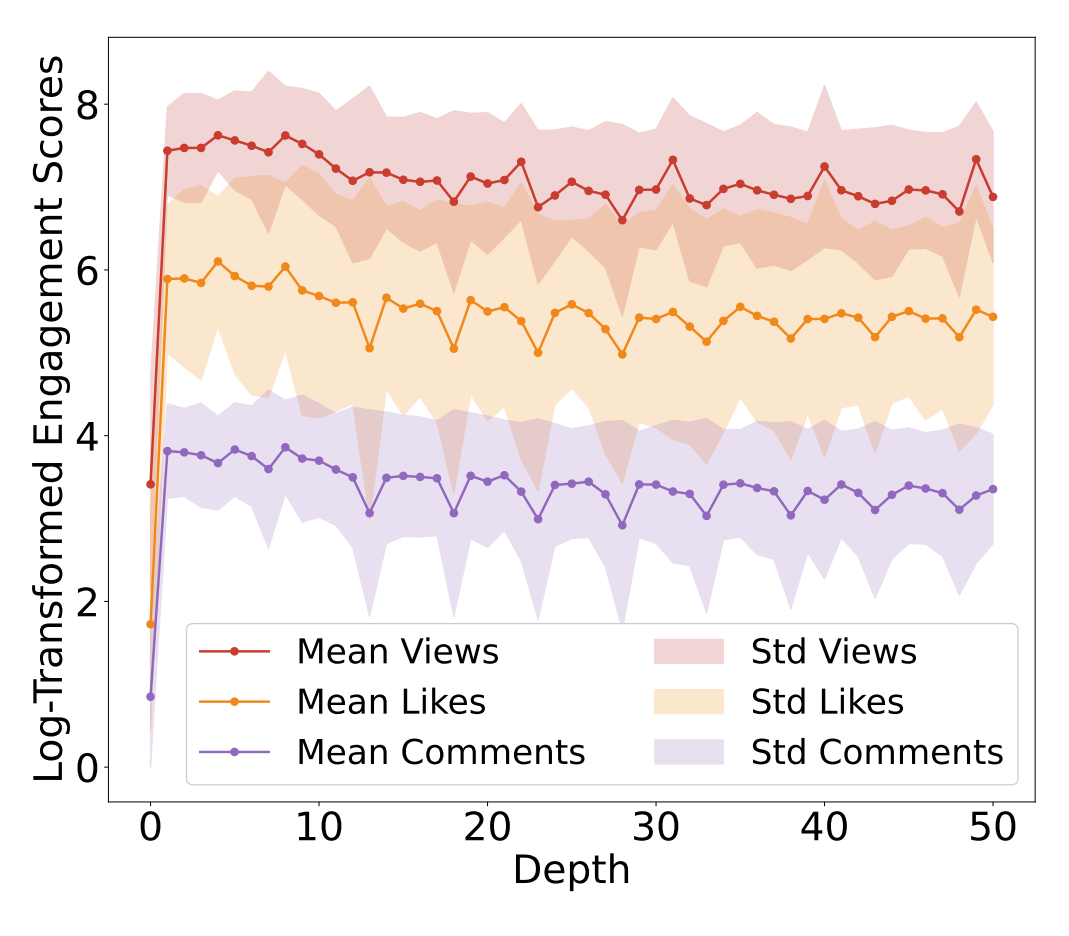}
         \caption{3-Second Watch Time}
         \label{engagement-3}
     \end{subfigure}
     \hfill
     \begin{subfigure}[b]{0.45\textwidth}
         \centering
         \includegraphics[width=\textwidth]{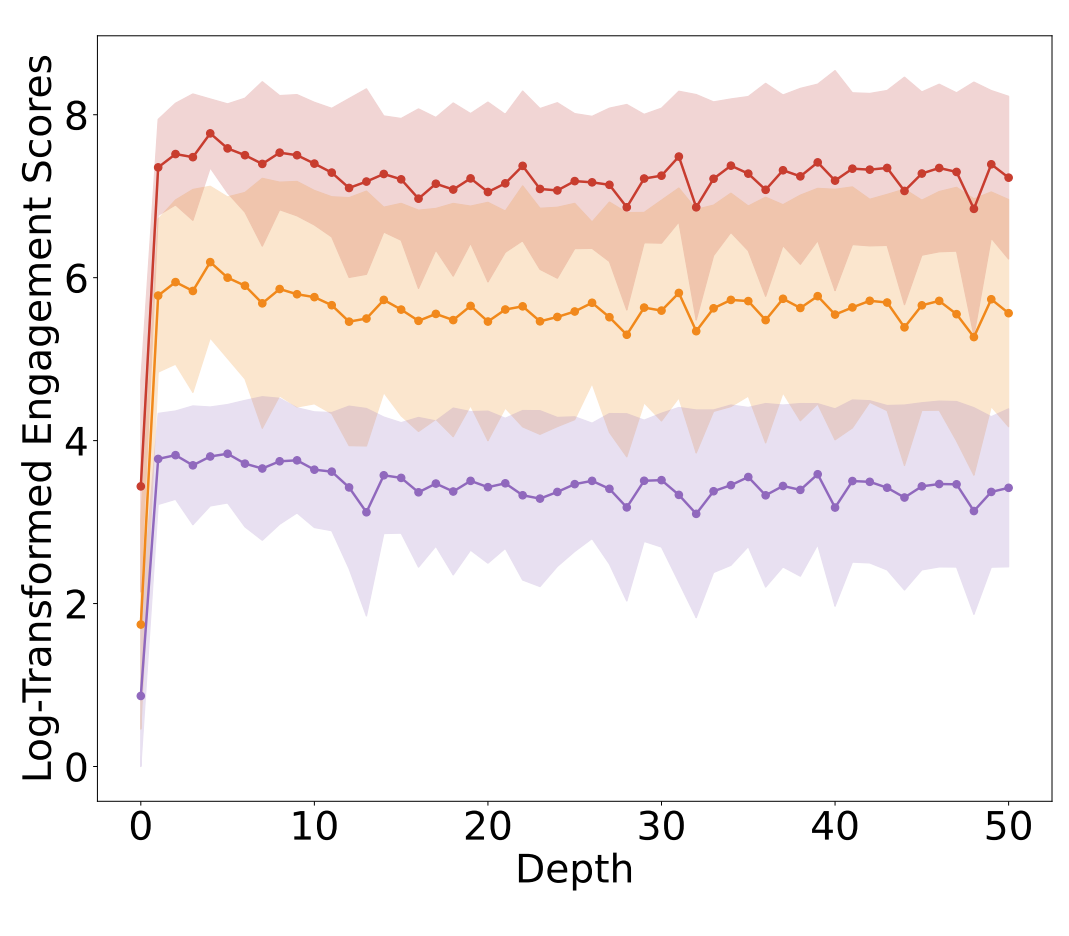}
         \caption{15-Second Watch Time}
         \label{engagement-15}
     \end{subfigure}
     \begin{subfigure}[b]{0.45\textwidth}
         \centering
         \includegraphics[width=\textwidth]{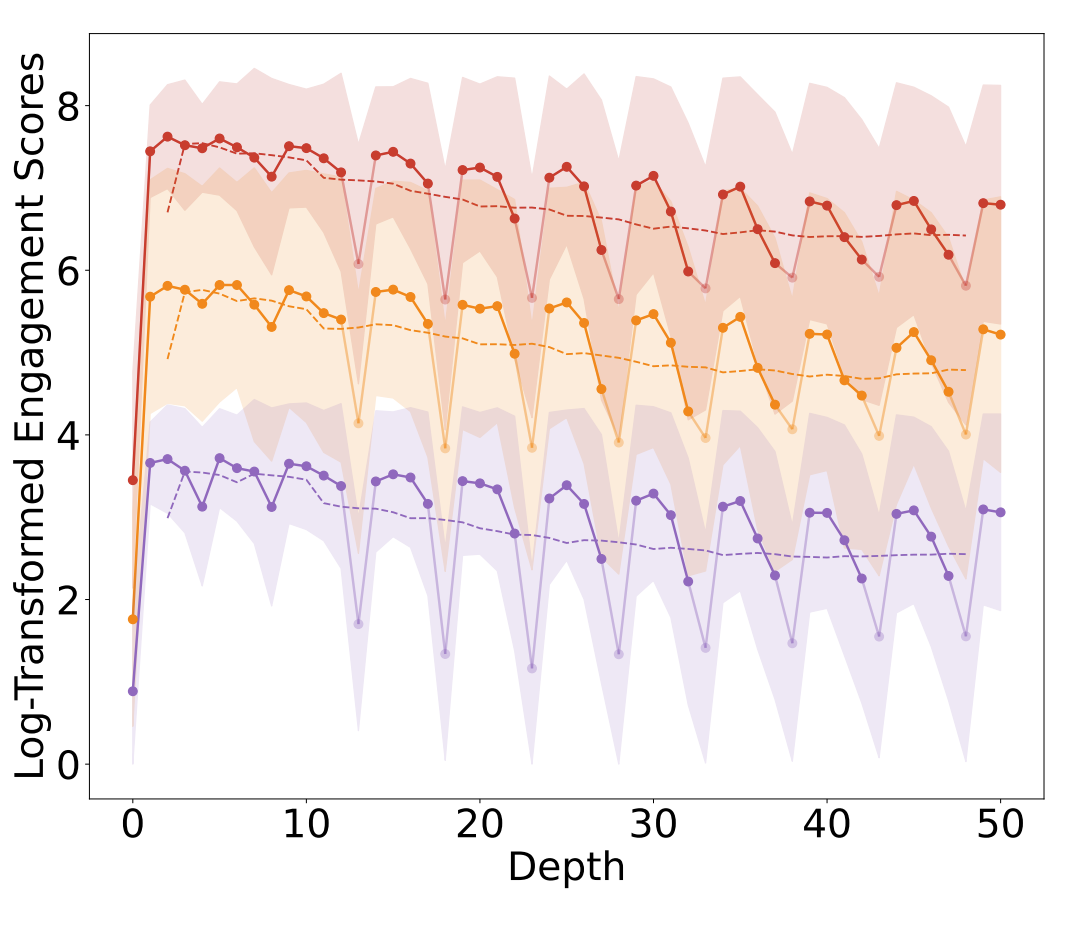}
         \caption{60-Second Watch Time}
         \label{engagement-60}
     \end{subfigure}
     \hfill
        \caption{Engagement Metrics of Short-Form Videos}
        \label{engagement-short}
\end{figure}

When comparing Figure \ref{engagement-short} with Figure \ref{engagement-long}, we can observe that both show an increase in the engagement metrics as the depths increase. However, the stark difference between the two sets of results is the sharp increase in engagement in the recommended YouTube Shorts videos in comparison to the regular videos. This shows that the YouTube Shorts recommendation algorithm more heavily prefers to recommend videos with high engagement compared the algorithm for recommending regular videos. 

Another thing to note is that the depth in Figure \ref{engagement-short} went up to 50, while the depth in Figure \ref{engagement-long} went only up to 5 depths. This was a decision based on the logistical difficulty in collecting more depths with exponentially increasing recommendations. Nevertheless, even looking at the first 5 depths in Figure \ref{engagement-short}, it becomes clear that there was a much more sudden change in the amount of engagements in the videos that were being recommended. 

Comparing the shift in the engagement metrics throughout the depths between the various wait times, it can be observed that different patterns emerged in each and that these patterns were caused by different reasons. Initially, the 3-second wait time communicates to the algorithm that the user did not find the content engaging and skipped the short-form video without watching it. This causes some spikes and variability in the engagement metrics as the algorithm tries to find something that the user will be interested in. This is most apparent when looking at the mean view line in Figure \ref{engagement-short} section (a). There were small spikes that were random and did not follow any pattern. 

When looking at the engagement values of the 15-second wait time, however, most of those random spikes disappear, and it can be observed that the line is a bit smoother. The largest difference between the engagement trend out of these three wait times shows up when analyzing the 60-second wait time engagement metrics. There seems to be a significant dip in engagement in exactly one in every five videos. When we observed this phenomenon occurring, we looked into the data more closely and realized that when the algorithm detects the user will watch anything they are recommended, YouTube's algorithm starts to recommend advertisements to the user to watch every fifth video. This was then confirmed by further manual testing, and it was verified that when watching YouTube Shorts, if the user fully watches every video, they will be periodically be shown ads, with this periodicity most commonly set at 5. 

\subsection{Emotion Analysis}

The engagement metrics of a video and its emotional response can be intertwined within the recommendation preference of the algorithm. It is important to measure both to see if there are independent correlations between these metrics. In this section, we take a look at the results for the emotional analysis for both short-form and long-form videos. Furthermore, we highlight the differences, if any, between the emotional metrics for the recommendations given by YouTube's short-form video recommendation algorithm and long-form recommendation algorithm. 

Before drawing conclusions, it is important to note that the emotional assessment resulted in a score between zero and one for each emotion. This was the same for both the analysis of short-form and of long-form videos. However, the emotional analysis for short-form videos was conducted on the metadata of videos, which included titles and descriptions. However, the long-form videos had extra data, such as the transcription and comments. We collected that data and ran it through the emotional analysis to see if there were any differences. Furthermore, when we tried to collect the transcripts for short-form videos manually, we saw that most of the short-form content that was collected had music only and no speech to transcribe and analyze. Since collecting and analyzing the transcripts of the few short-form videos that had speech would have caused skewed results, the emotional analysis of short-form videos were completed once using a combination of the title and the description. The YouTube Shorts emotionality analysis was split into the 3-second, 15-second, and 60-second wait times, identical to the split done in the analysis for engagement metrics.

\begin{figure}[h]
     \centering
     \captionsetup[subfigure]{justification=centering}
     \begin{subfigure}[b]{0.45\textwidth}
         \centering
         \includegraphics[width=\textwidth]{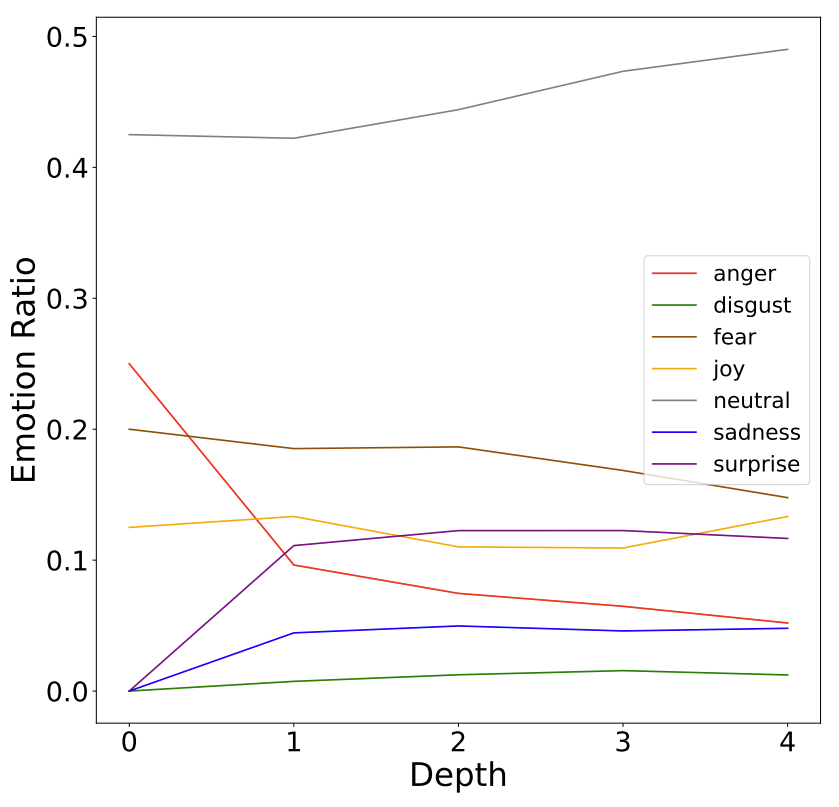}
         \caption{Emotions in the Title}
         \label{emotion-title}
     \end{subfigure}
     \hfill
     \begin{subfigure}[b]{0.45\textwidth}
         \centering
         \includegraphics[width=\textwidth]{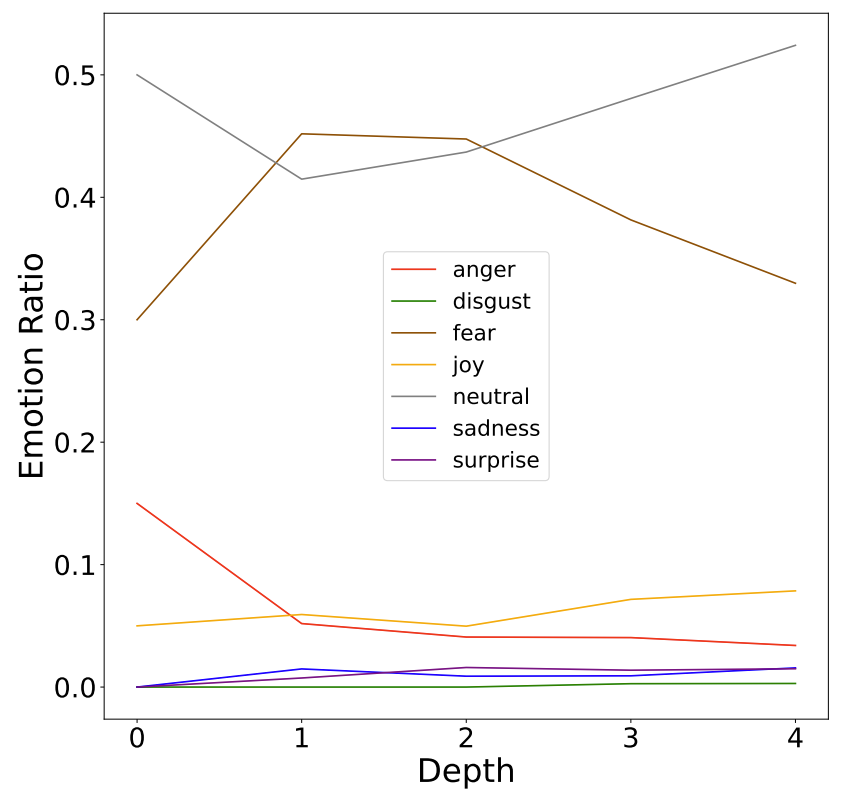}
         \caption{Emotions in the Description}
         \label{emotion-dascription}
     \end{subfigure}
     \begin{subfigure}[b]{0.45\textwidth}
         \centering
         \includegraphics[width=\textwidth]{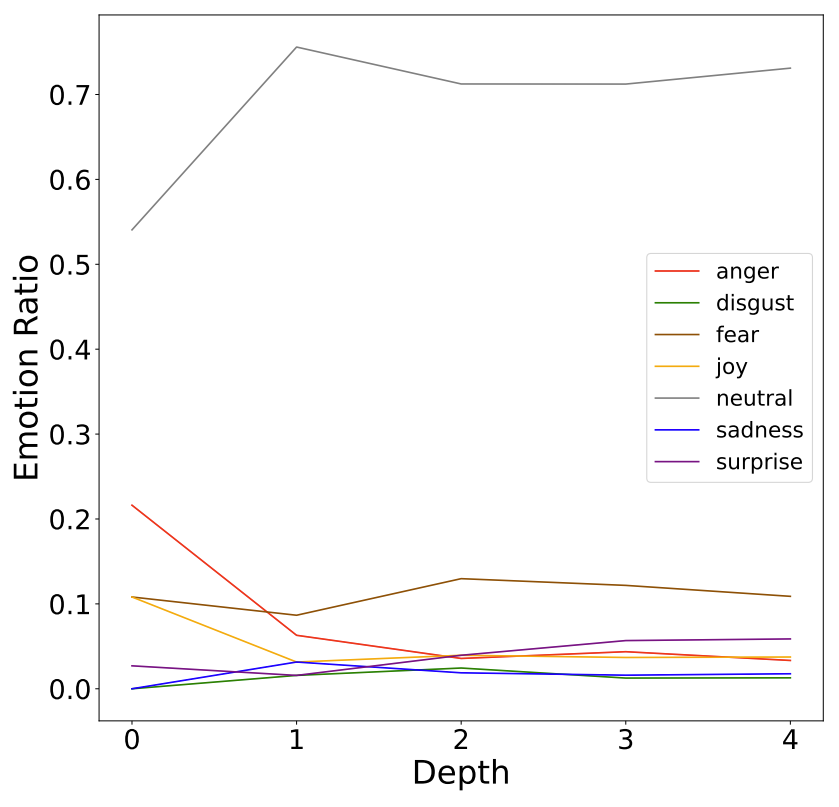}
         \caption{Emotions in the Transcription}
         \label{emotion-transcription}
     \end{subfigure}
     \hfill
     \begin{subfigure}[b]{0.45\textwidth}
         \centering
         \includegraphics[width=\textwidth]{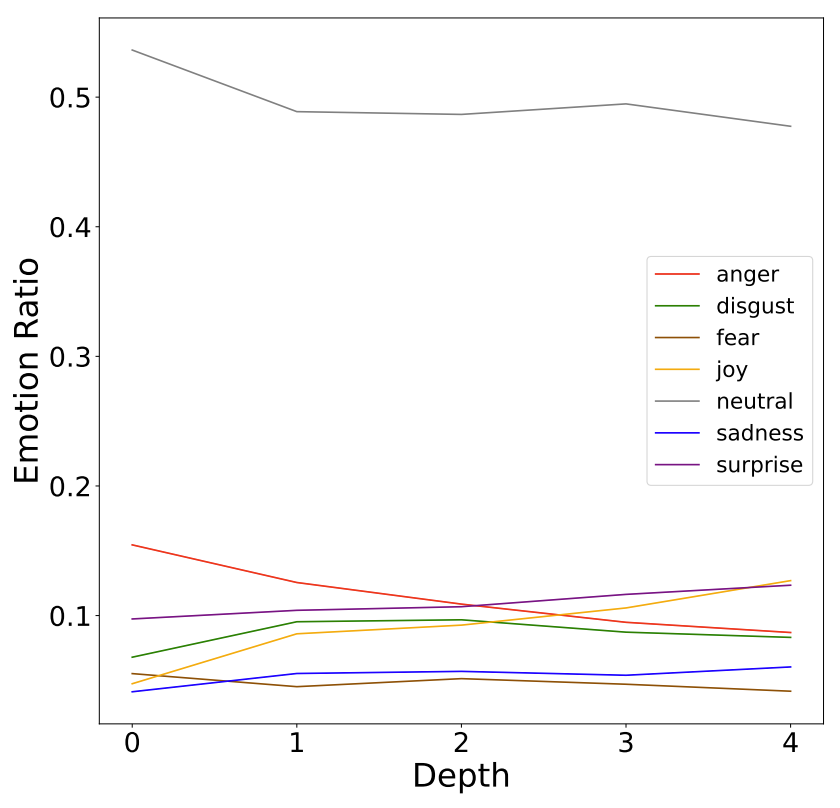}
         \caption{Emotions in the Comments}
         \label{emotion-comments}
     \end{subfigure}
        \caption{Emotional Analysis on Long-Form Videos}
        \label{emotion-long}
\end{figure}

In Figure \ref{emotion-long}, the emotions detected in the four different pieces of data extracted from the videos are shown as they change across the depths of recommendations. It is important to note that there was an emotion labeled as neutral that represents the absence of any clear emotion in the text provided to the model. The neutral emotion shows a clear increase when looking at the title, the description, and the transcription. Even when looking at the comments, there was only a slight decrease. This shows that the recommendation algorithm prefers to show content with increasingly neutral emotion and with a comments section with decreasing neutral emotion. Hence, although the algorithm recommends videos with lower amounts of clearly identifiable emotion, the emotional response from the users in the comments section seems to increase as the depths of the recommendation increases. 

Next, the most obviously consistent trend in emotional shifts through the depths was the decrease in the anger emotion in all four pieces of video text data extracted. This is by far the clearest piece of inference that can be drawn from Figure \ref{emotion-long}. Another emotional trend that was consistent throughout the sub-figures was the increase in surprise. These observations together are evidence that the algorithm actively tries to decrease the topics that incite anger in users and to recommend content that surprises users and keeps them engaged. This could be problematic if it means that the algorithm is suppressing critical information about current events that may rightfully cause the public to get angry, such as the escalation in military tension in the south-east region of China near Taiwan. 

Observing some of the other emotional trends, there is no clear inference that can be drawn from the trend in fear since there is a drastic difference between how the trends through the depths for the four different pieces of data. However, if the anomalies are ignored, especially looking at the title and the comments, there is a clear decrease in fear. Similarly, it is difficult to draw any conclusions about disgust or sadness since they both started low, ended low, and had no clear trends in the data. The emotion of joy had more subtle trends. When looking at the overall data extracted from the videos themselves, there was a gradual increase in joy. However, joy had a significant and clear increase when looking at the comments. Therefore, the recommendation algorithm prefers to recommend videos that incited joy in users.

\begin{figure}[h]
     \centering
     \captionsetup[subfigure]{justification=centering}
     \begin{subfigure}[b]{0.45\textwidth}
         \centering
         \includegraphics[width=\textwidth]{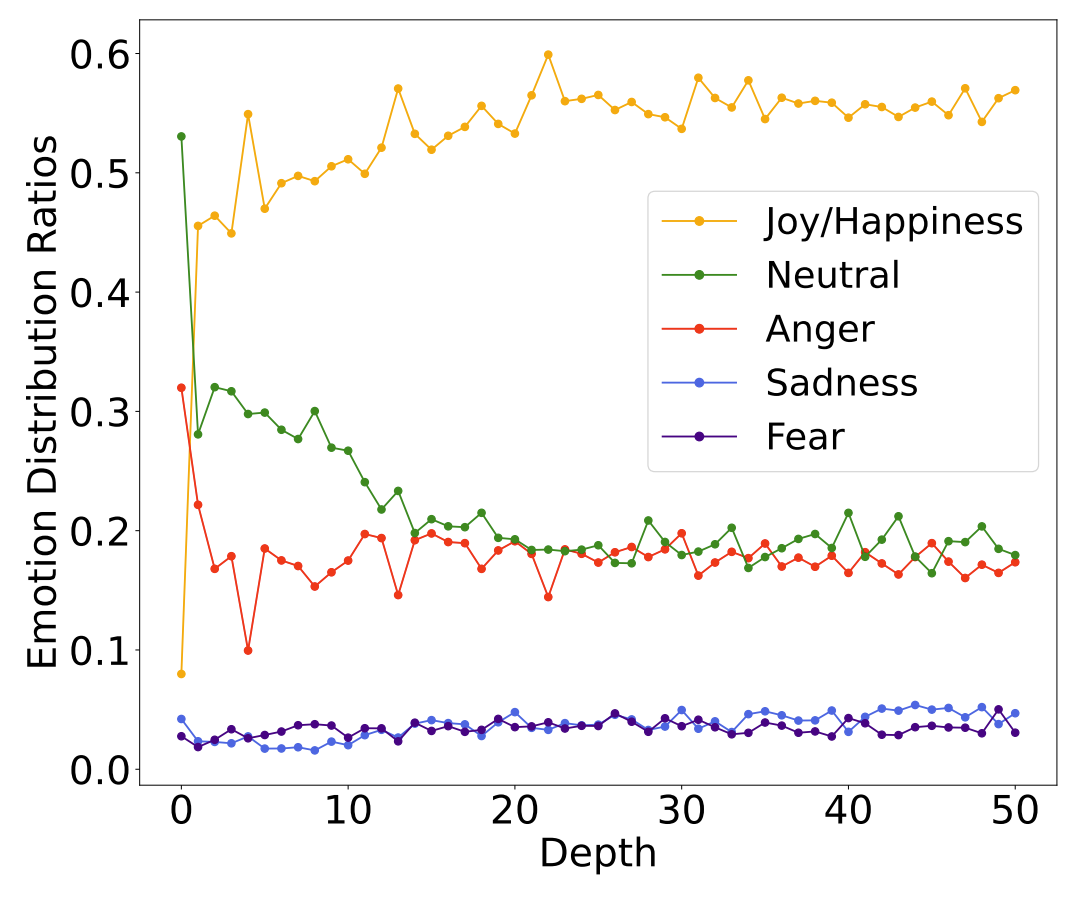}
         \caption{3-Second Watch Time}
         \label{emotion-3}
     \end{subfigure}
     \hfill
     \begin{subfigure}[b]{0.45\textwidth}
         \centering
         \includegraphics[width=\textwidth]{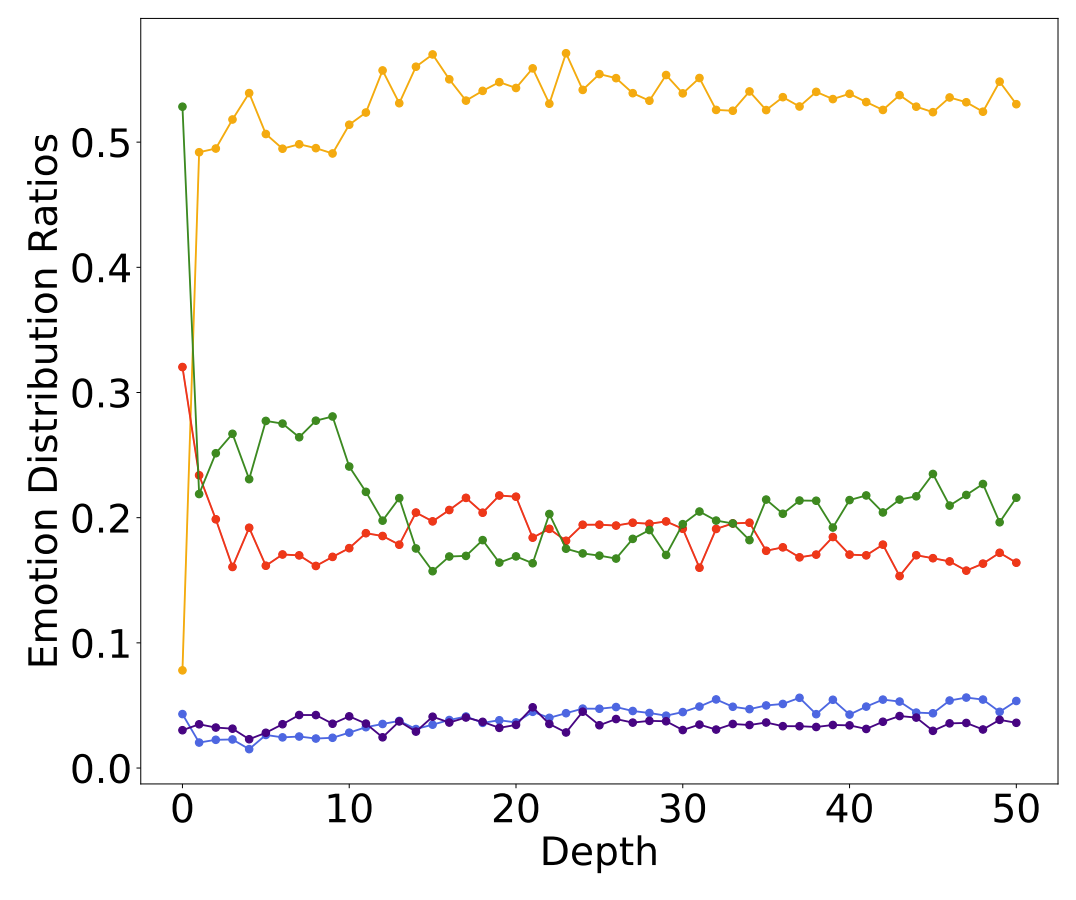}
         \caption{15-Second Watch Time}
         \label{emotion-15}
     \end{subfigure}
     \begin{subfigure}[b]{0.45\textwidth}
         \centering
         \includegraphics[width=\textwidth]{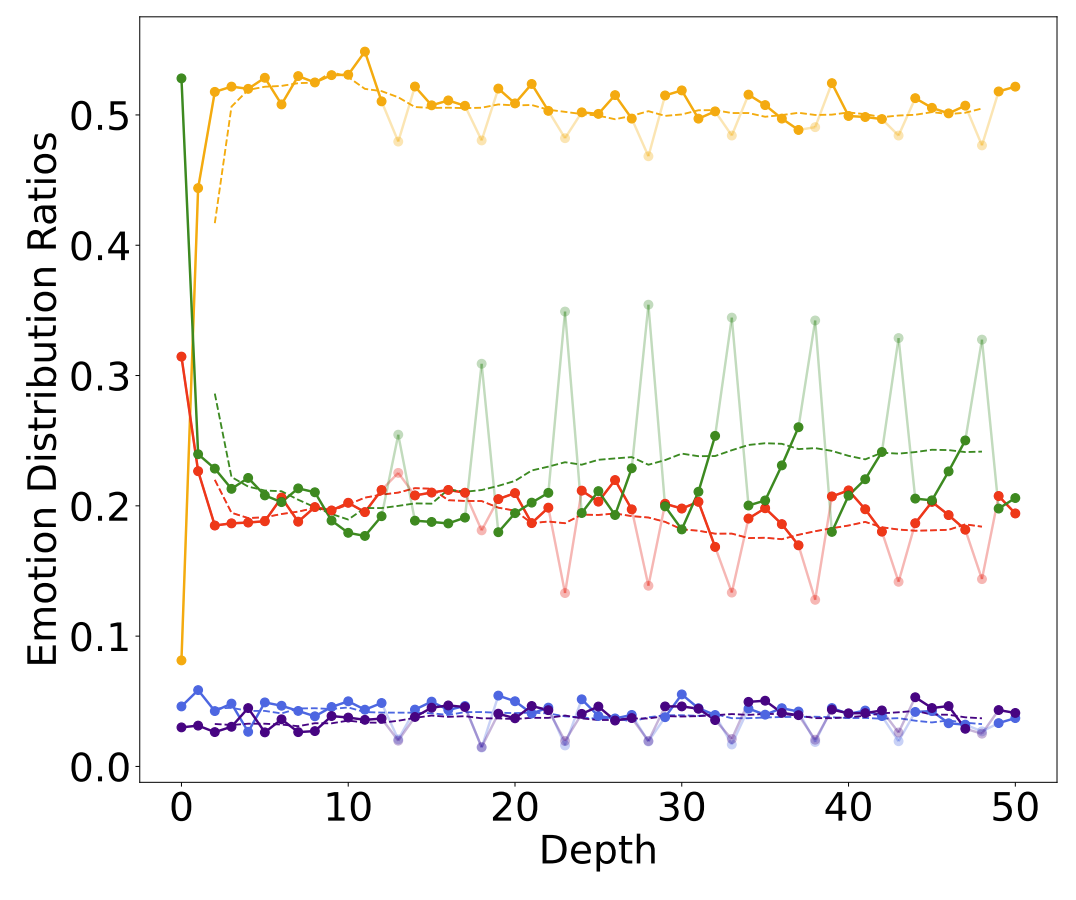}
         \caption{60-Second Watch Time}
         \label{emotion-60}
     \end{subfigure}
     \hfill
        \caption{Emotional Analysis on Short-Form Videos}
        \label{emotion-short}
\end{figure}

Now, looking at the emotional shifts in the recommendations for short-form content in Figure \ref{emotion-short}, it is immediately evident that the emotional trends happened in a more sudden fashion, starting from the first recommendation. There was an increase in joy in all watch times, similar to the recommendations for long-form videos. Contrary to the emotional shifts in long-form videos, there was a decrease in neutral emotion as joy increased. This was most likely caused by the fact that not only did joy increase a lot faster compared to regular videos but that the magnitude of this increase was also significantly higher, almost by a factor of five. 

Another similarity between the recommendation algorithms for regular YouTube videos versus YouTube Shorts was that they both reduce anger over time. Looking at all three sub-figures in Figure \ref{emotion-short}, anger has a clear decrease from its starting point around 0.3, and it later on tapered off right under 0.2. The emotional trends of sadness and fear were inconsistent and insignificant, hence, there was no clear conclusion that can be drawn from their results. 

One anomaly to point out for the emotional trends of the recommendations for YouTube Shorts was the spikes found in the 60-second watch time results. This, similar to the spikes in engagement values, was caused by the ads shown to users when their watch times were consistently high. When an advertisement was shown, anger spiked down due to the non-angering nature of advertisements, and this caused a resulting spike increase in neutral emotion.

\newpage

\subsection{Toxicity Analysis}

Toxicity analysis, although it might sound similar to emotionality analysis, focuses primarily on the aspects of social media that constitute harassment, divisiveness, or other forms of serious negativity. This type of behavior did not, and will generally not, show distinctly in categorical emotionality analysis. Furthermore, toxicity can draw attention, be infectious, and therefore be interconnected with the results of engagement. Our hypothesis going into this analysis was that the algorithm would prefer to recommend increasingly toxic content to the users. In this section, we have outlined the toxicity analysis results and drawn decisive conclusions from them. 

\begin{figure}[h]
     \centering
     \captionsetup[subfigure]{justification=centering}
     \begin{subfigure}[b]{0.45\textwidth}
         \centering
         \includegraphics[width=\textwidth]{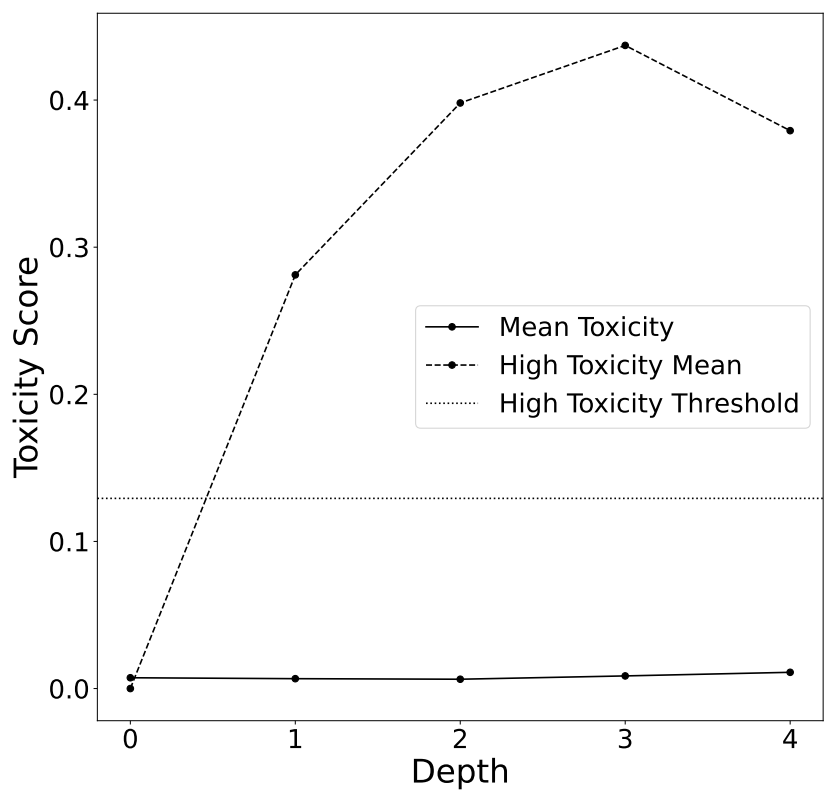}
         \caption{Toxicity in the Title}
         \label{toxicity-title}
     \end{subfigure}
     \hfill
     \begin{subfigure}[b]{0.45\textwidth}
         \centering
         \includegraphics[width=\textwidth]{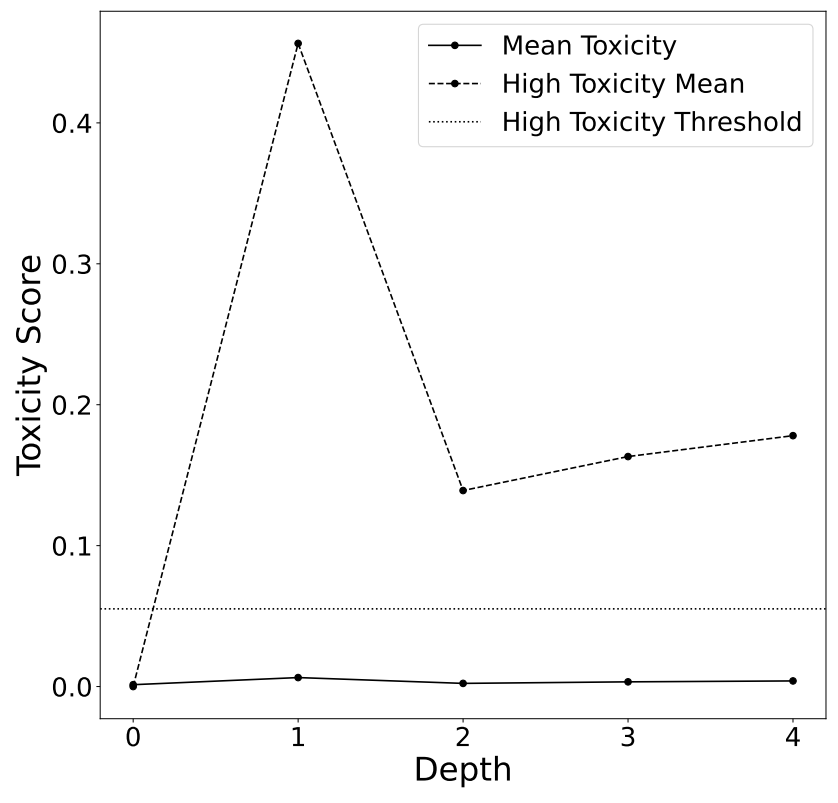}
         \caption{Toxicity in the Description}
         \label{toxicity-dascription}
     \end{subfigure}
     \begin{subfigure}[b]{0.45\textwidth}
         \centering
         \includegraphics[width=\textwidth]{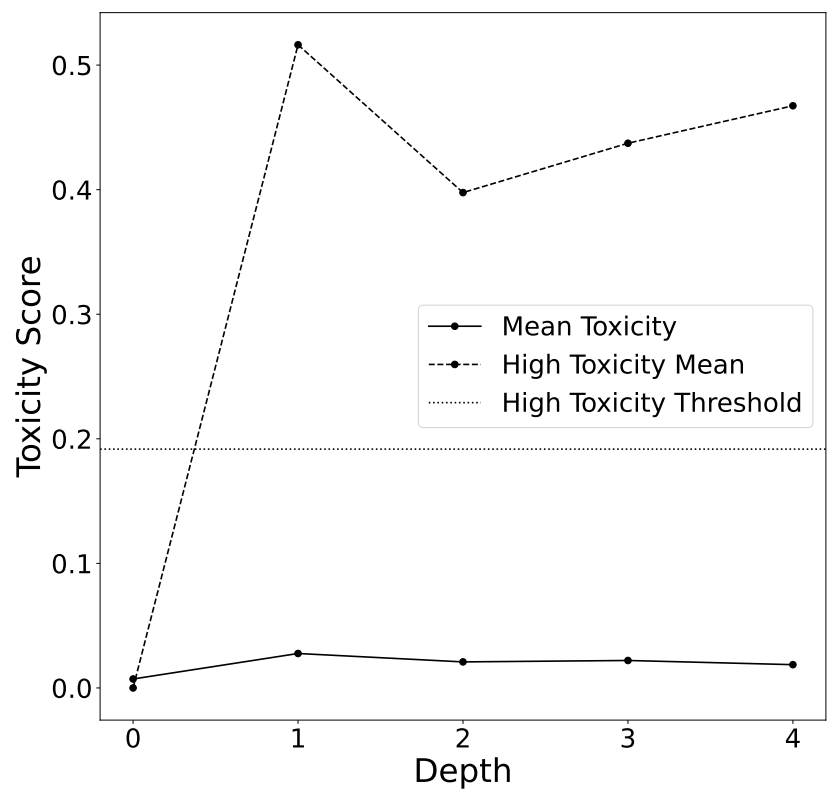}
         \caption{Toxicity in the Transcription}
         \label{toxicity-transcription}
     \end{subfigure}
     \hfill
     \begin{subfigure}[b]{0.45\textwidth}
         \centering
         \includegraphics[width=\textwidth]{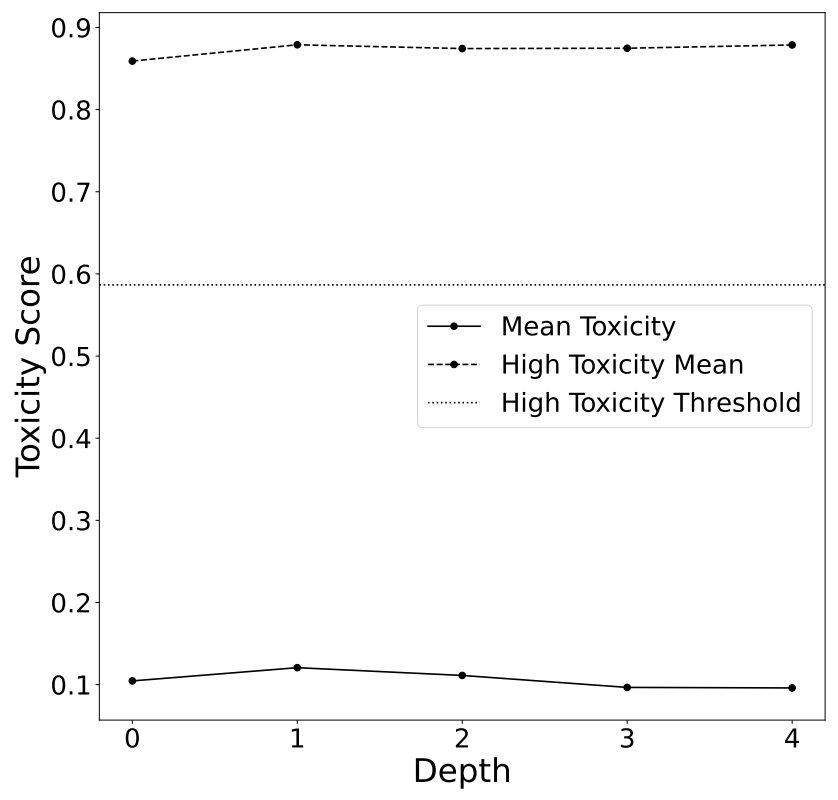}
         \caption{Toxicity in the Comments}
         \label{toxicity-comments}
     \end{subfigure}
        \caption{Toxicity Analysis on Long-Form Videos}
        \label{toxicity-long}
\end{figure}

After analyzing the results of the toxicity levels across the long-form recommendation depths in Figure \ref{toxicity-long}, it is immediately noticeable that the mean toxicity of all the videos—both the root videos and the recommended ones—was so low that it seems to be zero. However, when looking at the mean toxicity of the more toxic videos, there was a clear increase in the toxicity in the data extracted from the videos themselves (i.e. the title, description, and transcription). On the other hand, there was a consistently high amount of toxicity found in the video's comments, regardless of the depth of recommendation. These results show that the comments of YouTube videos were almost always toxic. Furthermore, the recommendation algorithm, as hypothesized, preferred to show increasingly toxic videos, irregardless of its possible effect on the users, in order to maximize engagement. 

\newpage

\begin{figure}[h]
     \centering
     \captionsetup[subfigure]{justification=centering}
     \begin{subfigure}[b]{0.49\textwidth}
         \centering
         \includegraphics[width=\textwidth]{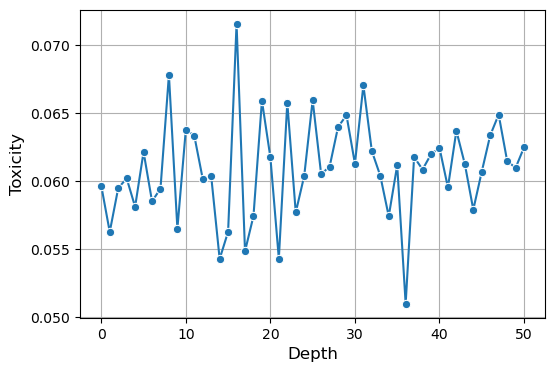}
         \caption{3-Second Watch Time}
         \label{toxicity-3}
     \end{subfigure}
     \hfill
     \begin{subfigure}[b]{0.49\textwidth}
         \centering
         \includegraphics[width=\textwidth]{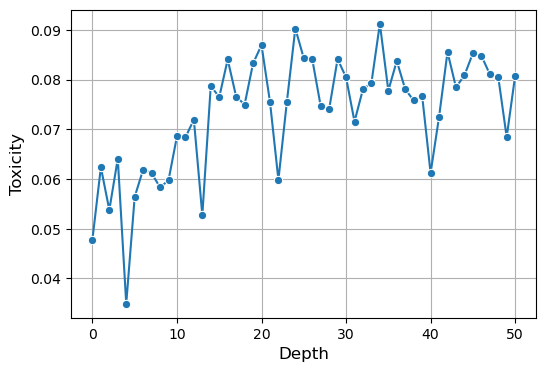}
         \caption{15-Second Watch Time}
         \label{toxicity-15}
     \end{subfigure}
     \begin{subfigure}[b]{0.49\textwidth}
         \centering
         \includegraphics[width=\textwidth]{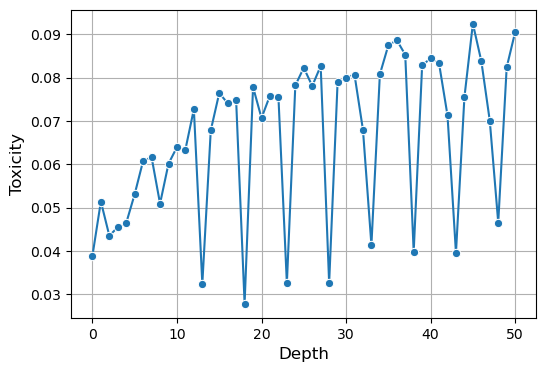}
         \caption{60-Second Watch Time}
         \label{toxicity-60}
     \end{subfigure}
     \hfill
        \caption{Toxicity Analysis on Short-Form Videos}
        \label{toxicity-short}
\end{figure}

Looking at the results of the toxicity analysis in the YouTube Shorts recommendations in Figure \ref{toxicity-short}, it is immediately evident that, when considering all three charts as representing the three different wait times, there was a consistent increase in toxicity throughout the recommendation depths. This, however, is the mean toxicity. When taking into consideration the scale of the results, by paying close attention to the values in the y-axis, it becomes clear that the average toxicity of all of the videos was close to zero. Since these charts are more closely zoomed in, the increase becomes more noticeable. This is similarly caused by the fact that there are outliers in the recommendations with much higher than average toxicity. This shows the willingness of the algorithm to recommend toxic videos to increase user engagement. 

Furthermore, focusing on the sub-figure Figure \ref{toxicity-60}, we saw that the advertisement spikes in the previous two results of the previous analyses shows up here as well. It can be inferred from these results that the toxicity of the advertisements was significantly lower than the videos the recommendation algorithm would otherwise recommend. 

\newpage


\section{Conclusion and Discussion} \label{conclusion}

In this research paper, we have successfully showed that collecting recommendation data on both regular YouTube videos and YouTube Shorts was possible. Not only did we develop a method that made this possible, but we improved upon it to make it practically viable enough to collect data for generating decisive results and analysis. This shows that, even if there are some roadblocks in the path, it is possible to devise unconventional solutions to overcome those challenges.

Furthermore, the timed results of multiprocessing applications, when compared to the base data collection methods for both long- and short-form content, shows that multiprocessing can be used to improve the processing time of light computational tasks by almost a factor equaling the number of processes used. The results further show that these improvements taper off at a certain point, and that that taper off point can be further improved upon by implementing clever solutions to reduce the computational demand of individual processes. 

This research also successfully compares the recommendation algorithms of regular YouTube videos and YouTube Shorts. The overarching conclusion of the analysis is that both of the recommendation algorithms push the users toward more engaging videos. Moreover, both algorithms preferred less angry and more joyful videos in their recommendations. They both also tended to recommend content with higher toxicity, seemingly to increase engagement. Additionally, the recommendation algorithm for YouTube Shorts recommended higher engaging videos much earlier and stronger than the algorithm for regular YouTube videos did. This faster deviation to high engagement videos—combined with the fact that recommendation in short-form content is a much stronger factor in deciding what users are exposed to—makes the recommendation algorithm for short-form content much more dangerous and biased. 

There were many successful conclusions drawn from the data, and there were also a few lessons learned. One of these lessons was to more carefully filter out advertisements when collecting recommendation data in the future. Nevertheless, this oversight helped show the engagement, emotion, and toxicity levels affected by YouTube Shorts advertisements. It also showed us that advertisements show up periodically and, more specifically, once every five videos. 

After conducting this study, it was also noticed that reacting with the content, such as liking, disliking, or commenting, could affect the recommendations. Hence, this research will be extended in the future by reacting positively with relevant content and reacting negatively with irrelevant content to measure if the algorithm picks up on user engagement. This study could additionally be extended by performing similar data collection and analysis methods on different platforms with short-form content to see if the recommendation algorithm in those platforms also behaves in a similar fashion. 

\newpage

\section*{Data Availability Statement}
The datasets analyzed during this study are available from the corresponding author upon reasonable request.

\section*{Statement of Disclosure}
On behalf of all authors, the corresponding author states that there is no conflict of interest.

\section*{Acknowledgments}

This research is funded in part by the U.S. National Science Foundation (OIA-1946391, OIA-1920920), U.S. Office of the Under Secretary of Defense for Research and Engineering (FA9550-22-1-0332), U.S. Army Research Office (W911NF-23-1-0011, W911NF-24-1-0078), U.S. Office of Naval Research (N00014-21-1-2121, N00014-21-1-2765, N00014-22-1-2318), U.S. Air Force Research Laboratory, U.S. Defense Advanced Research Projects Agency, Arkansas Research Alliance, the Jerry L. Maulden/Entergy Endowment at the University of Arkansas at Little Rock, and the Australian Department of Defense Strategic Policy Grants Program. Any opinions, findings, and conclusions or recommendations expressed in this material are those of the authors and do not necessarily reflect the views of the funding organizations. The researchers gratefully acknowledge the support. 

\bibliography{sn-bibliography}


\end{document}